\documentclass[reprint,amsmath,amssymb,aps,prb,]{revtex4-1}
\usepackage{graphicx}% Include figure files
\usepackage{empheq}
\usepackage{dcolumn}% Align table columns on decimal point
\usepackage{bm}% bold math
\usepackage{amsmath}
\usepackage[dvipsnames]{xcolor}
\usepackage{relsize}
\usepackage{lipsum}
\usepackage{subfigure}
\newenvironment{rightcases}
  {\left.\begin{aligned}}
  {\end{aligned}\right\rbrace}
  
  \DeclareMathOperator{\sign}{sign}
  
\begin{document}
\preprint{APS/123-QED}

\title{Modeling superconductivity in the background of a spin-vortex checkerboard}

\author{Anastasia V. Aristova}
\email{anastasia.aristova@skolkovotech.ru}
\author{ $\!\! ^{\dagger}$ Vivek K. Bhartiya}
\thanks{A. V. Aristova and V. K. Bhartiya  contributed equally to this work.}
 \author{Boris V. Fine}
 \email{b.fine@skoltech.ru}
\affiliation{Skolkovo Institute of Science and Technology, Nobel Str. 3, 143026 Moscow, Russia}
\date{6 August, 2019}

\begin{abstract}
We introduce a microscopic model aimed at describing the behavior of fermionic excitations in the background of a magnetic texture called ``spin-vortex checkerboard". This texture was proposed previously as a possible alternative to stripes to interpret the experimental phenomenology of spin and charge modulations in 1/8-doped lanthanum cuprates. The model involves two kinds of interacting fermionic excitations residing in spin-rich and spin-poor  regions of the modulated structure. It is a generalization of another model developed earlier for the so-called ``grid checkerboard". The principal terms of our model describe the decay of fermionic pairs belonging to spin-poor regions into single fermions occupying spin-rich regions and vice versa. These terms induce intricate fermionic correlations throughout the system but fall short of inducing superconductivity unless arbitrarily small hopping terms are added to the model Hamiltonian. We present the mean-field solution of the model, including, in particular, the temperature dependence of the energy gap. The latter is  found to be in a good overall agreement with available experimental data for high-$T_c$ cuprate superconductors.  
\end{abstract}

\pacs{TBD}

\maketitle

\section{Introduction}
Interplay between superconductivity and the onset of electronic spin and charge modulations in cuprate superconductors remains one of the intriguing and unresolved issues in the field of high-temperature superconductivity. Manifestations of electronic modulations are reported in a broad doping range for several families of cuprates --- most noticeably around the doping level of 1/8\cite{TraquadaStripe1995,Yamada1998,Hoffman2002,McElroy2003,Vershinin2004,Hanaguri2004,Abbamonte2005,McElroy2005,Komiya2005,Wise2008,
Edurado2014,Comin2015,Comin2016,peng2018re}. 

\begin{figure}
  \label{fig:spinbackground}
  \centering
    \subfigure[]{\includegraphics[scale=0.27]{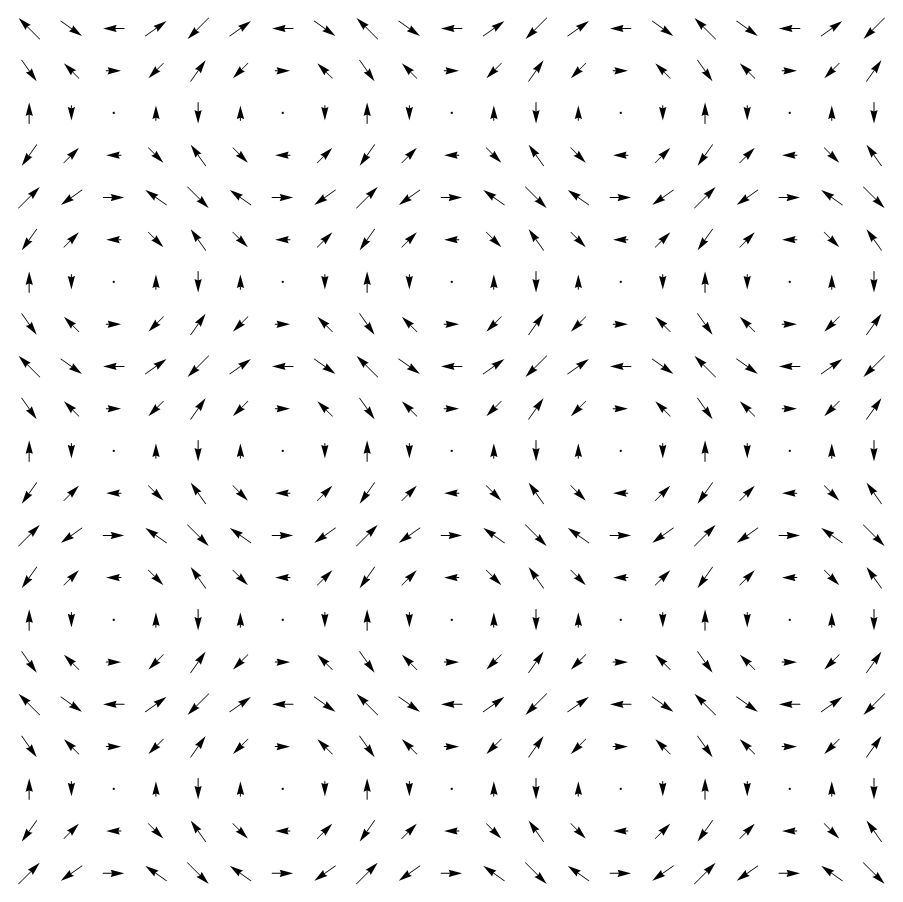}\label{fig:Checkers}}\quad 
    \subfigure[]{\includegraphics[scale=0.56]{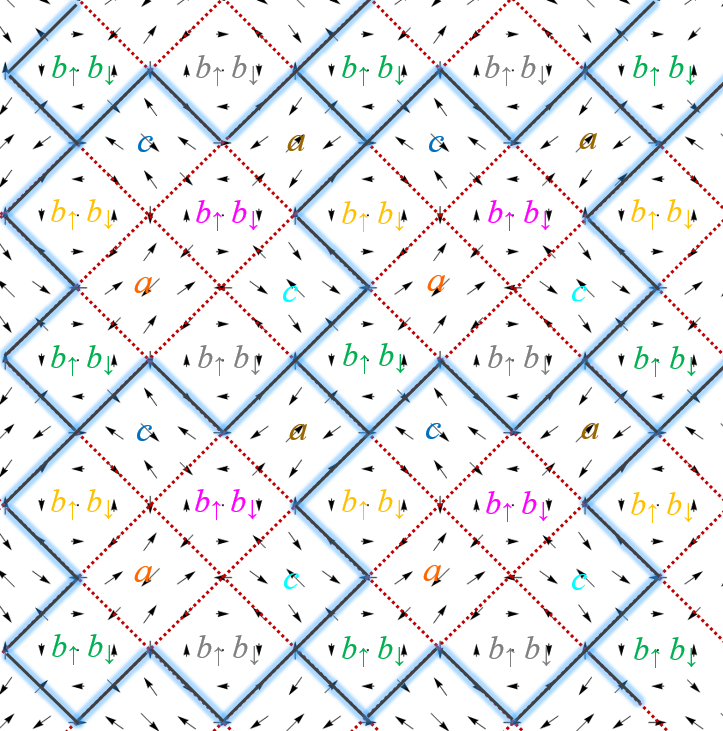}\label{fig:Cunitcell}}
  \caption{(Color online). (a) Spin-vortex checkerboard. Each arrow represents average spin polarization on a square lattice formed by Cu atoms within CuO$_2$ planes. (b) Partition of the spin-vortex checkerboard into plaquets associated with {\it a}-, {\it b}-, and {\it c}-states. Thick lines indicate the borders of  unit cells of the modulated structure. Each unit cell (also shown in Fig.\ref{fig:unitcell}) includes  two {\it a}-states, two {\it c}-states and four {\it b}-states.}
\end{figure}

For 1/8-doped lanthanum cuprates, the modulated structure is widely believed to exhibit one-dimensional pattern often referred to as ``stripes" \cite{TraquadaStripe1995,Yamada1998}. Yet the principal aspects of the same experimental evidence  are also consistent with the possibility of two-dimensional modulations called ``checkerboards" \cite{Tranquada1999,Mitsen2004,Robertson2006,SchriefferBook2007,Boris2004,Boris2007,Boris2011,Boris2013}. 
The experiment-based arguments discriminating between stripes and checkerboards in 1/8-doped lanthanum cuprates are, at present, rather indirect. At the same time, the issue cannot be resolved on purely theoretical grounds, because it requires accuracy of the calculations of the ground-state energy not achievable by first-principles theories. 
A particularly focused effort to investigate the consequences of the checkerboard scenario was made in  Ref.\cite{Boris2004}.
That analysis was based on a particular kind of checkerboard called ``grid". Later, the grid checkerboard was shown to be inconsistent with the results of spin-polarized neutron scattering experiment of Ref.\cite{Christensen2007}. This experiment, however, did not rule out another version of a checkerboard representing a two-dimensional arrangement of spin vortices \cite{Boris2007}  shown in Fig.\ref{fig:Checkers}. Somewhat similar noncollinear spin textures were also considered in Refs. \cite{Seibold1998,Berciu1999,Timm2000,Koizumi2008,Wilson2009,Azzouz2010}. Recently an analogous superstructure called "spin-vortex crystal" was proposed to exist in iron-based superconductors\cite{Avci2014,Bohmer2015,Ohalloran2017,Meier2017}.

Spin-vortex checkerboard in the context of cuprates was introduced in Ref.\cite{Boris2007} . Its various properties  were analyzed in Refs.\cite{Boris2007,Boris2011,Boris2013,Dolgirev2017}.  So far, however, this analysis has not touched the superconducting properties. In Section~\ref{microscopic} of the present article, we introduce a microscopic model aimed at describing superconductivity in the background of spin-vortex checkerboard. The model is a generalization of another one proposed in Ref.\cite{Boris2004} for the grid checkerboard. In Sections~\ref{solution} and \ref{solution2}, we find the mean-field solution of the generalized model in the same way as done in Ref.\cite{Boris2004}. However, our subsequent analysis in Section~\ref{superconducting} differs from that of Ref.\cite{Boris2004} in one significant respect, namely, we find that both the original model and its generalization made in Section~\ref{microscopic} do not induce non-zero superfluid stiffness. (In this regard, we correct an error made in Ref.\cite{Boris2004}.) We further show in Section~\ref{sec:stiff} that non-zero superfluid stiffness emerges once arbitrarily small hopping terms are included in the modeling.   

 Experimentally, the onset of static spin modulations in 1/8-doped lanthanum cuprates --- spin vortices or stripes --- largely suppresses three-dimensional superconductivity but appears to coexist with two-dimensional superconductivity \cite{Li2007,Berg2007,Tranquada2008,li2018tuning}. In Section~\ref{scenarios}, we show that, similar to the stripe scenario,  the suppression of the three dimensional superconductivity in spin-vortex scenario can be explained by the displacement of the modulation patterns in the adjacent CuO$_2$ planes. 
 
 As further argued in Section~\ref{scenarios}, the generic setting in cuprates beyond 1/8-doped lanthanum family,  possibly, involves the fluctuating counterpart of the static spin-vortex texture used in the present work. These fluctuations are likely caused by system's proximity to the threshold of electronic phase separation \cite{Fine2008}. They  are expected to couple spin, charge and lattice degrees of freedom  --- see e.g.\cite{Egami2010}. In this respect, our model illustrates the potential of the general two-component scenarios\cite{Ranninger2010,Pawlowski2010} in the limit of initially localized components for describing the superconductivity in cuprates. In such scenarios, the first component represents unpaired fermions, while the second component represents preformed fermionic pairs.
 The spin-vortex checkerboard just suggests us the coupling connectivity between the low-energy fermionic states. We assume that in other cuprate families, the inhomogeneous patterns have a two-dimensional character, checkerboard-like or more disordered, where regions of stronger and weaker antiferromagnetic correlations alternate with other, but the size of these regions may be different for different dopings, and the doping dependence of that size may be different for different cuprate families. This would be consistent with rather diverse phenomenology summarized in Ref.\cite{Comin2016}. The dimensionality of charge modulations was, in particular, discussed for the yttrium family of curates in Refs.\cite{comin2015broken,fine2016comment,comin2016comment}.
 
 Finally, in Section~\ref{experiment}, we compare the temperature evolution of the energy gap obtained from the mean-field solution of our model with experiments in bismuth cuprates. Several important but technical calculations behind the reported results are placed in the Appendices.

    \begin{figure}
     \includegraphics[width=0.4\textwidth]{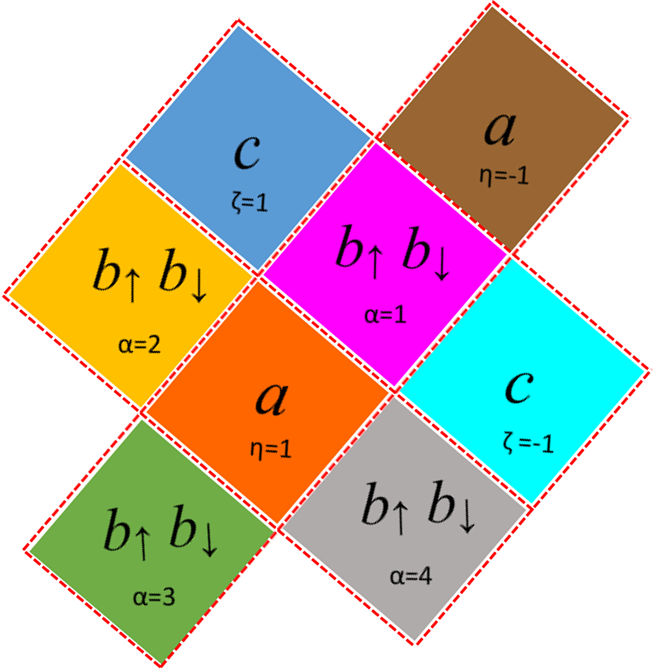}
\caption{(Color online). Unit cell from Fig. \ref{fig:Cunitcell} with labels $\alpha$, $\eta$ and $\zeta$ as introduced in the text. Colors represent different quasiparticle states as follows:  {\it b}-states with  $\alpha=1$ (pink), $\alpha=2$ (yellow), $\alpha=3$ (green), and $\alpha=4$ (gray);    even {\it a}-state [$\eta = 1$] (orange), odd {\it a}-state [$\eta = -1$]  (brown); even {\it c}-state [$\zeta = 1$] (blue), odd {\it c}-state [$\zeta = -1$]  (cyan).}\label{fig:unitcell}
  \end{figure}

\section{Microscopic model: principal terms}
\label{microscopic}

The model to be considered has two different kinds of fermionic states physically located in magnetic and non-magnetic parts of the underlying spin texture.  The general reasoning for constructing the model is the same as in Ref.\cite{Boris2004}. Namely, the entire texture is divided into plaquets having different kinds of spin background, and then, for each plaquet, only one-particle fermionic states closest to the chemical potential are retained. Given that plaquets are rather small, it can be estimated\cite{Boris2004} that the spacing of one-particle energies within each plaquet is of the order of 40~meV, which implies that, for temperatures much smaller than 400K, it is appropriate to retain only the levels closest to the chemical potential.

As shown in Fig.\ref{fig:Cunitcell},  spin vortex checkerboard can be represented as a square superlattice of $8d\times 8d$, where $d$ is the period of the underlying square lattice of Cu atoms. We denote the total number of such unit cells in the system as $N$. Each unit cell in Fig.\ref{fig:Cunitcell} is further divided into four spin-polarized plaquets and four spin unpolarized plaquets.
We expect that the lowest one-particle states in spin-polarized  plaquets are non-spin-degenerate, and hence we include exactly one state per plaquet. We refer to two of the resulting states as ``{\it a}-states" and to the remaining two as ``{\it c}-states". Two different kinds of {\it a}-states are distinguished by index $\eta= \pm 1$, and {\it c}-states --- by index  $\zeta= \pm 1$. Two {\it a}-states or two {\it c}-states with different values of $\eta$ or $\zeta$ respectively are expected to have orthogonal spin wave functions that can be obtained from each other by spin inversion. The lowest-energy states of spin-unpolarized plaquets around the cores of spin vortices are assumed to be spin-degenerate. We, therefore, place two fermionic states on each such plaquet with spins ``up" or ``down" along any chosen direction. We call them ``{\it b}-states". Since the spin texture contains four nonequivalent kinds of spin-vortex cores, we distinguish the corresponding {\it b}-states by index $\alpha = 1,2,3,4$ and by spin index $\uparrow$ or $\downarrow$ --- see Fig. \ref{fig:unitcell}.

We now construct a low-energy Hamiltonian similar to that of Ref.\cite{Boris2004}. We expect that charge carriers in the background of spin-vortex checkerboard are heavily dressed by antiferromagnetic fluctuations, which should lead to relatively large effective masses. Charge carriers are also supposed to experience strong Coulomb repulsion, which should further prevent them from hopping. Therefore, in the zeroth order, we neglect single-fermion hopping between checkerboard plaquets. (Hopping will be treated in Section~\ref{sec:stiff} as a small correction.) As far as the interaction terms are concerned, we use the same "heavy-mass/Coulomb-repulsion" argument to include in the model only those terms that do not change  the center-of-mass/center-of-charge positions of the two fermions involved. 
We also neglect a possible "on-site" coupling of two $b$-states on the same plaquet --- in our preliminary analysis, this term would not change the essential features of the solution, but we still plan to investigate it elsewhere. The above assumptions amount to a relatively crude overall approximation, which should nevertheless allow us to capture the principal behavior of the variational solution and, at the same time, avoid introducing too many adjustable parameters.  

We are, finally, left with terms representing on-site energies $\epsilon_a$, $\epsilon_b$ and $\epsilon_c$ (with $\epsilon_a = \epsilon_c$) and with the following two-kinds of effective interaction terms, namely: two {\it a}-states or two {\it c}-states adjacent to a given spin vortex core making transitions to the two {\it b}-states inside the core or vice versa. The resulting Hamiltonian is: 
\begin{multline}\label{eq:originalHamlt}
H = \sum_{i,\eta}\epsilon_{a}  a^{\dagger}_{i\eta} a_{i\eta}+  \sum_{i,\zeta}\epsilon_{c} c^{\dagger}_{i\zeta} c_{i\zeta} +   \sum_{i,\alpha,\eta}\epsilon_{b}b^{\dagger}_{i\alpha \eta} b_{i\alpha \eta}\\ + g \sum_{i,\alpha} \left[ \left( b^{\dagger}_{i \alpha \uparrow}b^{\dagger}_{i \alpha \downarrow} a_{j \text{e}_{[i,\alpha]}} a_{k \text{o}_{[i,\alpha]}}+h.c\right)\right.\\ \left. + \left( b^{\dagger}_{i \alpha \uparrow}b^{\dagger}_{i \alpha \downarrow} c_{m \text{e}_{[i,\alpha]}} c_{n \text{o}_{[i,\alpha]}}+h.c\right) \right],
   \end{multline}
where $g$ is the interaction constant, $\epsilon_a$, $\epsilon_b$ and $\epsilon_c$ are on-site energies defined with respect to the chemical potential $\mu$, which we set equal to zero, index $i$ labels  unit cells depicted in Fig \ref{fig:unitcell}, and indices $\eta$, $\alpha$ and $\zeta$ label the plaquets within the unit cell as illustrated in Fig \ref{fig:unitcell}. Following Ref.\cite{Boris2004}, whenever the specific value of subscripts $\eta$ or $\zeta$ is fixed, as is the case in the interaction term of Hamiltonian (\ref{eq:originalHamlt}), we use subscript ``e" for $\eta = 1$ or $\zeta = 1$ referring to the corresponding plaquets as ``even" and subscript ``o" for $\eta = -1$ or $\zeta = -1$ referring to the corresponding plaquets as ``odd". Double-subscripts notations such as $a_{j \text{e}_{[i,\alpha]}}$ imply that {\it a}-states labeled as $\{j \text{e}\}$ must be adjacent to the {\it b}-states labeled as $\{ i,\alpha \}$.

If all terms containing {\it c}-states are removed from Hamiltonian (\ref{eq:originalHamlt}), the result would be exactly equivalent to the Hamiltonian considered in Ref \cite{Boris2004}. Since {\it c}-states do not directly couple to {\it a}-states, and since {\it c}-states have the same connectivity with the {\it b}-states as {\it a}-states (but shifted),  the mean-field solutions of the two models are very similar  with the only difference being that {\it b}-states now experience mean field from both {\it a}-states and {\it c}-states, which, in turn, makes that mean field two times larger, and, as a result, the value of the superconducting transition temperature becomes modified.

Since the entire mean-field solution has nearly the same structure and logic  as that of Ref. \cite{Boris2004}, below we only include the formal structure of the derivation and the results, leaving the justification mostly to Ref. \cite{Boris2004}. 

As explained later in Section~\ref{sec:stiff}, the system described by Hamiltonian~(\ref{eq:originalHamlt}) will not be superconducting, because it will have zero superfluid stiffness. In order to make it superconducting,  arbitrarily small hopping terms will need to be added. However, the model based on Hamiltonian~(\ref{eq:originalHamlt}) is easier to handle, and its solution already captures the important aspects of the resulting superconducting phase, such as the energy gap. 
 
\section{Bogoliubov transformations}
\label{solution}
In the model considered, each of the fermionic states couples to relatively few other states, which makes a mean-field solution rather approximate. We, nevertheless, assume that it gives at least the right qualitative picture of model's behavior. The first step of this solution is to introduce the Bogoliubov transformation for {\it b}-states within the same plaquet:
\begin{eqnarray}\label{eq:bgvtransb}
\begin{aligned}
   b_{i\alpha \uparrow} &= s B_{i\alpha \uparrow} + w e^{i\varphi_{\alpha}} B^{\dagger}_{i\alpha \downarrow}, \\
   b_{i\alpha \downarrow} &= s B_{i\alpha \downarrow} - w e^{i\varphi_{\alpha}} B^{\dagger}_{i\alpha \uparrow},
   \end{aligned}
   \end{eqnarray}
     where $s$ and $w$ are positive real numbers satisfying a constraint arising from canonical fermionic anticommutation relations     
   \begin{equation}\label{eq:bBog}
   s^{2} +w^{2} =1,
   \end{equation}
and $\varphi_{\alpha}$ are the transformation phases, which are to be determined later by minimizing system's energy.
   
  Substituting Bogoliubov transformation  for {\it b}-states   in (\ref{eq:originalHamlt}) and keeping only the thermal averages of terms that do not change the occupations of Bogoliubov $B$-states, we obtain partially averaged Hamiltonian,
    \begin{multline}\label{eq:Avbstate}
H_{a}= 8 \epsilon_{b} N \left[ s^{2} n_{B} + w^{2} ( 1- n_{B}) \right]\\+ \epsilon_{a} \sum_{i,\eta} a^{\dagger}_{i\eta} a_{i\eta}+ \epsilon_{c} \sum_{i,\zeta} c^{\dagger}_{i\zeta} c_{i\zeta}\\+gsw ( 1- 2n_{B}) \mathlarger{\mathlarger{\sum}}_{\alpha} 
  \bigg[ (e^{-i \varphi_{\alpha}} \sum_{ij} a_{j \text{e}_{[i,\alpha]}} a_{k \text{o}_{[i,\alpha]}} +h.c )\\ + (e^{-i \varphi_{\alpha}} \sum_{ij} c_{m \text{e}_{[i,\alpha]}} c_{n \text{o}_{[i,\alpha]}} +h.c )\bigg], 
\end{multline}
where $\epsilon_{B}$ is the energy of $B$-quasiparticles and
\begin{equation}
n_{B} = \frac{1}{\exp{\frac{\epsilon_{B}}{T}}+1},
\label{nB}
\end{equation}
is their occupation number. We set $k_{{\text B}}=1$.

As explained in Ref.\cite{Boris2004}, in order to assure proper fermionic anticommutation relations for the Bogoliubov counterparts of {\it a}- and {\it c}-states,  the Bogoliubov transformation for these states should be made in the quasimomentum space.  
Therefore, we need to rewrite the Hamiltonian in terms of the real-space Fourier transforms for {\it a}- and {\it c}-operators. To do the Fourier transforms, we first change the notations from $a_{i \eta}$, $c_{i \eta}$ to the notations $a(\mathbf{r})$, $c(\mathbf{r})$, where $\mathbf{r}$ indicates the position of the center of the respective plaquet. We also need to define the following vectors --- all in units of underlying crystal lattice period $d$: 
\begin{equation}
\mathbf{L} = (0,4),
\label{L}
\end{equation}
which connects an even {\it a}-state with an adjacent even {\it c}-state, and 
    \begin{eqnarray}
    \begin{aligned}
    \mathbf{R_{1}} &= (4,4), &\\
    \mathbf{R_{2}} &= (-4,4),& \\
    \mathbf{R_{3}} &= (-4,-4), &\\
    \mathbf{R_{4}} &= (4,-4), &
    \end{aligned}
    \end{eqnarray}
which connect an even {\it a}-state with four adjacent odd {\it a}-states. The subscript $\alpha$ in ${\bf R}_{\alpha}$ is chosen such that $\alpha$th {\it b-}states are located between the pairs {\it a}-states connected by vector ${\bf R}_{\alpha}$ originated from an even {\it a}-state.
Now, we define the position of each unit cell by the position ${\bf r}_\text {e}$ of an even {\it a}-state within this cell. Therefore, even {\it a}-states are located at a set of positions $\{ {\bf r}_\text{e} \}$, odd {\it a}-states at $\{ {\bf r}_\text{e}  + {\bf R}_{1} \}$, even {\it c}-states at $\{ {\bf r}_\text{e}  + {\bf L} \}$, and odd {\it c}-states at $\{{\bf r}_\text{e} + {\bf L}  + {\bf R}_1 \}$.
Finally, we rewrite the Hamiltonian (\ref{eq:Avbstate}) as follows, 
%%%%%%%%%%%%%%%%%%%%%%%%%%%%%%%%%%%%%%
\begin{multline}\label{eq:evenodd}
H_{a} = 8 \epsilon_{b} N \left[ s^{2} n_{B} + w^{2} ( 1- n_{B}) \right] 
\\
+  \sum_{\mathbf{r}_\text{e}} \left\{ 
   \epsilon_{a} a^{\dagger}({\bf r}_\text{e}) a({\bf r}_\text{e}) 
   + \epsilon_{a}  a^{\dagger}( {\bf r}_\text{e}  + {\bf R}_{1}) a( {\bf r}_\text{e}  + {\bf R}_{1})
   \right.\\ \left.
   +\epsilon_{c}   c^{\dagger}({\bf r}_\text{e}  + {\bf L}) c({\bf r}_\text{e}  + {\bf L}) \right.\\ \left.
   +\epsilon_{c}  c^{\dagger}({\bf r}_\text{e} + {\bf L}  + {\bf R}_{1}) c({\bf r}_\text{e} + {\bf L}  + {\bf R}_{1})
   \right\}
   \\ 
   + gsw ( 1- 2n_{B})\sum_{\alpha} \bigg[ 
   \left(e^{-i \varphi_{\alpha}} \sum_{{\bf r}_\text{e}} a({\bf r}_\text{e}) a({\bf r}_\text{e} + {\bf R}_{\alpha}) +h.c \right)  \\ + \left(e^{-i \varphi_{\alpha}} \sum_{{\bf r}_\text{e}} c({\bf r}_\text{e} +{\bf L}) c({\bf r}_\text{e}+{\bf L} +{\bf R}_{5-\alpha}) +h.c \right)\bigg],
   \end{multline}
   %%%%%%%%%%%%%%%%%%%%%%%%%%%%%%%%%%%%%%
We now explicitly write separate Fourier transforms for even and odd {\it a}- and {\it c}-states as follows:
\begin{align}\label{eq:FT}
a_\text{e}(\mathbf{k})& = \sqrt{\frac{1}{N}} \sum_{{\bf r}_\text{e}} a({\bf r}_\text{e}) e^{-i {\bf k r}_\text{e}},  \\
a_\text{o}(\mathbf{k})& = \sqrt{\frac{1}{N}} \sum_{{\bf r}_\text{e}} a({\bf r}_\text{e}+{\bf R_{1}}) e^{-i {{\bf k}({\bf r}_\text{e}+{\bf R}_{1})}},  \\
c_\text{e}({\bf k})& = \sqrt{\frac{1}{N}} \sum_{{\bf r}_\text{e}} c({\bf r}_\text{e}+{\bf L}) e^{-i {\bf k} ({\bf r}_\text{e}+{\bf L})}, \\
c_\text{o}({\bf k})& = \sqrt{\frac{1}{N}} \sum_{{\bf r}_\text{e}} c({\bf r}_\text{e}+{\bf L}+{\bf R}_{1}) e^{-i {\bf k} ({\bf r}_\text{e}+{\bf L} + {\bf R}_{1})}.
\label{eq:FT4}
\end{align}
Since the superlattice periods for each of the above four kinds of states are the same,  the sets of wave vectors $\{ \mathbf{k}\}$ are also the same, even though the corresponding states are shifted with respect to each other in real space. Substituting these transformations to 
(\ref{eq:evenodd}), we obtain
%%%%%%%%%%%%%%%%%%%%%%%%%%%%%%%%%%%%%%%%%%%%%%%%%%%%
\begin{multline}\label{eq:FTac}
  H = 8 \epsilon_{b} N \left[ s^{2} n_{B} + w^{2} ( 1- n_{B}) \right]+ 
  \epsilon_{a} \sum_{\mathbf{k}} a^{\dagger}_\text{e}(\mathbf{k}) a_\text{e}(\mathbf{k}) +\\ \epsilon_{a} \sum_{\mathbf{k}} a^{\dagger}_\text{o}(\mathbf{k}) a_\text{o}(\mathbf{k})+\epsilon_{c} \sum_\mathbf{k} c^{\dagger}_\text{e}(\mathbf{k}) c_\text{e}(\mathbf{k})+\epsilon_{c} \sum_{\mathbf{k}} c^{\dagger}_\text{o}(\mathbf{k}) c_\text{o}(\mathbf{k})\\
  +gsw ( 1- 2n_{B})\sum_{\mathbf{k}} \bigg[
  \left( a_\text{e}(\mathbf{k}) a_\text{o}(\mathbf{-k})V(\mathbf{k})  +h.c \right)\\+  ( c_\text{e}(\mathbf{k}) c_\text{o}(\mathbf{-k})\tilde{V}(\mathbf{k})  +h.c )\bigg] ,
 \end{multline}
where
 %%%% expression for V(k) %%%%%%%%%%
   \begin{eqnarray} \label{eq:VK}
     V(\mathbf{k}) &=& \sum_{\alpha} \exp^{-i\varphi_{\alpha}-i \mathbf{k}\mathbf{R}_{\alpha}}, \\
    & = & 2 \exp \left[ -i \frac{\varphi_{1} + \varphi_{3}}{2} \right] \cos \left[\mathbf{k}\mathbf{R}_{1} + \frac{\varphi_{1} - \varphi_{3}}{2}\right]\nonumber\\&&
    +2 \exp \left[ -i \frac{\varphi_{2} + \varphi_{4}}{2} \right] \cos \left[\mathbf{k}\mathbf{R}_{2} + \frac{\varphi_{2} - \varphi_{4}}{2}\right],\nonumber
     \end{eqnarray}
      %%%% expression for  V`(k)%%%%%%%%%%%%%
  and
  \begin{eqnarray}\label{eq:VKT}
    \tilde{V}(\mathbf{k}) &=& \sum_{\alpha} \exp^{-i\varphi_{\alpha}-i \mathbf{k}\mathbf{R}_{\alpha-1}} \\
    & = & 2 \exp \left[ -i \frac{\varphi_{1} + \varphi_{3}}{2} \right] \cos \left[\mathbf{k}\mathbf{R}_{2} + \frac{\varphi_{3} - \varphi_{1}}{2}\right] \nonumber\\&&
    +2 \exp \left[ -i \frac{\varphi_{2} + \varphi_{4}}{2} \right] \cos \left[\mathbf{k}\mathbf{R}_{1} + \frac{\varphi_{4} - \varphi_{2}}{2}\right].\nonumber
  \end{eqnarray}
%%%%%%%%%%%%%%%%%%%%%%%%%%%%%%%%%%%%%%%%%%%%%%%%%%%%

Bogoliubov transformations for  {\it a}- and {\it c}-states can now be defined as,
\begin{align} 
%%%%%%%% a- states canonical transformation%%%%%
a_\text{e}(\mathbf{k}) & = u(\mathbf{k}) A_\text{e}(\mathbf{k}) + v(\mathbf{k}) e^{i\phi_{a}(\mathbf{k})} A_\text{o}^{\dagger}(\mathbf{-k}),\label{eq:BTaeststes} \\
a_\text{o}(\mathbf{-k}) &= u(\mathbf{k}) A_\text{o}(\mathbf{-k}) - v(\mathbf{k}) e^{i\phi_{a}(\mathbf{k})} A_\text{e}^{\dagger}(\mathbf{k}),\label{eq:BTaoststes}\\
%%%%% c- states canonical transformation%%%%
c_\text{e}(\mathbf{k}) & = p(\mathbf{k}) C_\text{e}(\mathbf{k}) + q(\mathbf{k}) e^{i\phi_{c}(\mathbf{k})} C_\text{o}^{\dagger}(\mathbf{-k}), \label{eq:BTceststes}\\
c_\text{o}(\mathbf{-k}) &= p(\mathbf{k}) C_\text{o}(\mathbf{-k}) - q(\mathbf{k}) e^{i\phi_{c}(\mathbf{k})} C_\text{e}^{\dagger}(\mathbf{k}),\label{eq:BTcoststes}
\end{align} 
where $u(\mathbf{k}),$  $v(\mathbf{k})$ and $p(\mathbf{k}),$  $q(\mathbf{k})$ are the real-valued  coefficients for {\it a}- and {\it c}-states respectively, subjected to a constraint arising from the fermionic canonical commutation relations for $A$- and $C$-operators:
\begin{align} \label{eq:uvpqcons}
u(k)^{2} +v(k)^{2}&=1,\\
p(k)^{2}+q(k)^{2}&=1,
\end{align}
and  $\phi_{a}$ and $\phi_{c}$ are complex phases --- all to be found by the energy minimization.

We now complete the following steps: (i) substituting the above canonical transformation for {\it a}- and {\it c}-states (\ref{eq:BTaeststes}-\ref{eq:BTcoststes}) into Hamiltonian (\ref{eq:FTac}),  then (ii) obtaining the energy of the system by summing over the thermal averages of the diagonal terms --- the result is given in Appendix~\ref{ap:total} by Eq.\eqref{eq:energywphase}---  and then (iii) minimizing the resulting energy with respect to the choice of phases $\phi_{a}(\mathbf{k})$ and $\phi_{c}(\mathbf{k})$. As explained in Appendix~\ref{ap:total}, these steps lead to conditions:
\begin{align}
\cos [ \phi_{V}(\mathbf{k})+ \phi_{a}(\mathbf{k})] &=1,\label{eq:phia+v=0} \\
\cos [\phi_{\tilde{V}}(\mathbf{k})+ \phi_{c}(\mathbf{k})]& =1,\label{eq:phic+v=0} 
\end{align}
were $\phi_{V}(\mathbf{k})$ and $\phi_{\tilde{V}}(\mathbf{k})$ are the complex phases of $V(\mathbf{k})$ and $\tilde{V}(\mathbf{k})$ respectively, which, in turn, depend on phases $\{\varphi_{\alpha} \}$. [Phases $\phi_{a}(\mathbf{k}$) and $\phi_{c}(\mathbf{k})$ do not need to be obtained explicitly, because they will not enter any quantity computed in this paper.] With the above conditions, the expression for the energy of the system becomes:
%%%%%%%%%%%%%%%%%%%%%%%%%%%%%%%%%%%%%%%
\begin{multline} \label{eq:energyfinal} 
 E =  8 \epsilon_{b} N \left[ s^{2} n_{B} + w^{2} (1- n_{B}) \right]  \\+ 2 \epsilon_{a} \sum_{\mathbf{k}} \left\lbrace u^{2}(\mathbf{k})n_{A}(\mathbf{k}) + v^{2}(\mathbf{k})[1-n_{A}(\mathbf{k})] \right\rbrace \\+2 \epsilon_{c} \sum_{\mathbf{k}} \left\lbrace p^{2}(\mathbf{k})n_{C}(\mathbf{k}) + q^{2}(\mathbf{k})[1-n_{C}(\mathbf{k})] \right\rbrace \\+ 2gsw \left( 1- 2n_{B} \right) 
   \sum_{\mathbf{k}} \bigg[ u(\mathbf{k})v(\mathbf{k})(1-2n_{A}(\mathbf{k})) |V(\mathbf{k})|\bigg. \\ \bigg.+ p(\mathbf{k})q(\mathbf{k})(1-2n_{C}(\mathbf{k}))]|\tilde{V}(\mathbf{k})| \bigg],
  \end{multline}
 where,
 \begin{equation}\label{eq:na}
 n_{A}(\mathbf{k}) =\frac{1}{\exp{\frac{\epsilon_{A}(\mathbf{k})}{T}}+1},
\end{equation}
\begin{equation}\label{eq:nc}
 n_{C}(\mathbf{k}) = \frac{1}{\exp{\frac{\epsilon_{C}(\mathbf{k})}{T}}+1},
 \end{equation}
are the Bogoliubov quasiparticle occupation number and $\epsilon_{A}(\mathbf{k})$, $\epsilon_{C}(\mathbf{k})$ are their energies obtained in the next section.
 
\section{Single-particle excitations, energy gap and the critical temperature}
\label{solution2}

As argued in Ref \cite{Boris2004}, the chemical potential of the system is, likely, to coincide with either  $\epsilon_{b}$ or  $\epsilon_{a}$ (same as $\epsilon_{c}$), which, given our convention $\mu = 0$, means that either $\epsilon_{b} = 0$ or $\epsilon_{a} = \epsilon_{c} = 0$. Below, we treat these two cases separately, referring to them as ``Case I" and ``Case II" respectively, and also refer to the case of $\epsilon_{a} = \epsilon_{b}=\epsilon_{c} = 0$ as ``critical". 

The coefficients of the Bogoliubov transformations for both Cases I and II are obtained in Appendices \ref{ap:caseI} and \ref{ap:caseII}, respectively, by minimizing the total energy (\ref{eq:energyfinal}) at fixed quasiparticle occupation numbers\cite{LP}. We then substitute those coefficients back to Eq (\ref{eq:energyfinal}) and obtain the energy of a Bogoliubov quasiparticle by taking derivative of the total energy (\ref{eq:energyfinal}) with respect to the quasiparticle occupation numbers   $n_{A}(\mathbf{k})$, $n_{C}(\mathbf{k})$ or $n_{B}$.

\subsection{Case I:   $\epsilon_{b}=0$ }

 In Case I, the above procedure gives the following quasiparticle energies: 
 \begin{eqnarray}\label{eq:EaenergyI}
\epsilon_{A}(\mathbf{k}) & =& \sqrt{\epsilon_{a}^{2} + \frac{1}{4} g^{2} (1-2n_{B})^{2} |V(\mathbf{k})|^{2}}, 
 \end{eqnarray}
  \begin{eqnarray} \label{eq:EcenergyI}
\epsilon_{C}(\mathbf{k}) & =& \sqrt{\epsilon_{c}^{2} + \frac{1}{4} g^{2} (1-2n_{B})^{2} |\tilde{V}(\mathbf{k})|^{2}}, 
\end{eqnarray}   
 \begin{multline}\label{eq:EbenergyI}
\epsilon_{B}
 =\frac{g^{2}}{16N}(1-2n_{B}) \sum_{\mathbf{k}}  \bigg[ \frac{(1-2n_{A}(\mathbf{k}))}{\epsilon_{A}(\mathbf{k})}|V(\mathbf{k})|^{2} \\+  \frac{(1-2n_{C}(\mathbf{k}))}{\epsilon_{C}(\mathbf{k})} |\tilde{V}(\mathbf{k})|^{2}\bigg].
 \end{multline}
 
 The mean-field approach now requires finding a nontrivial solution for  $\epsilon_{A}(\mathbf{k})$, $\epsilon_{C}(\mathbf{k})$, and $\epsilon_{B}$ from Eqs.(\ref{nB}, \ref{eq:na}, \ref{eq:nc},\ref{eq:EaenergyI}, \ref{eq:EcenergyI}, \ref{eq:EbenergyI}). In general, it can only be done numerically, but one can also obtain a closed analytical equation for the critical temperature $T_c$ using the fact that, near the transition, the ordered state is close to the normal states, which allows one to use the limits
 \begin{eqnarray*}
 \epsilon_{A}(\mathbf{k}) &\rightarrow &\epsilon_{a}, \\
 \epsilon_{C}(\mathbf{k}) &\rightarrow &\epsilon_{c}, \\
 (1-2n_{B}) &\rightarrow & \frac{\epsilon_{B}}{2T_{c}}.
 \end{eqnarray*}
 This gives
  \begin{equation}
  \begin{split}
 T_{c}   =  \frac{g^{2}}{8} \left[ 
 \left(\frac{\exp(|\epsilon_{a}|/T_{c})-1}{\exp(|\epsilon_{a}|/T_{c}) + 1} \right)\frac{1}{{|\epsilon_{a}|}} \right. \\ + \left.  \left(\frac{\exp(|\epsilon_{c}|/T_{c})-1}{\exp(|\epsilon_{c}|/T_{c})+1}\right)\frac{1}{{|\epsilon_{c}|}} \right],
 \end{split}
 \label{Tc1}
\end{equation}
 from which the mean-field $T_{c}$ can be obtained numerically. 
 
 As explained in Ref.\cite{Boris2004}, the energy densities of $A$- and $C$-states described by Eqs.~(\ref{eq:EaenergyI}) and (\ref{eq:EcenergyI}) have Van Hove singularities located in the both cases at the value 
 \begin{equation}
 \Delta = \sqrt{\epsilon_{a}^{2} + g^{2} (1-2n_{B})^{2}},
 \label{Delta1}
 \end{equation}
 corresponding to $|V(\mathbf{k})| = |\tilde{V}(\mathbf{k})| = 2$. We refer to $\Delta$ as the ``energy gap''.
 As $T \rightarrow T_c$, $\Delta$ approaches not zero but $|\epsilon_a|$, which we associate with the pseudogap.

\subsection{Case II:   $\epsilon_{a}=\epsilon_{c}=0$ }
 
 Following the same procedure as for Case I, we obtain: 

\begin{equation}\label{eq:EbenergyII}
  \epsilon_{B} = {\sqrt{\epsilon_{b}^{2}+\frac{g^{2}}{64}\tilde{C}^{2}}},
 \end{equation}
 
  \begin{eqnarray} \label{eq:EaenergyII}
  \epsilon_{A}(\mathbf{k}) &=
  &\frac{g^{2} (1-2n_{B})\tilde{C}|V(\mathbf{k})|}{16\epsilon_{B}},
  \end{eqnarray}
  \begin{eqnarray} \label{eq:EcenergyII}
   \epsilon_{C}(\mathbf{k}) &=
  &\frac{g^{2} (1-2n_{B})\tilde{C}|\tilde{V}(\mathbf{k})|}{16\epsilon_{B}},  
 \end{eqnarray}
  where  
  \begin{align} 
  \tilde{C} &=C_{a} +C_{c}, \label{eq:Ctilde}\\
   C_{a}&= \frac{1}{N}\sum_{\mathbf{k}}(1 - 2n_{A}(\mathbf{k})) |V(\mathbf{k})|,\label{eq:Ca}\\
   C_{c}&= \frac{1}{N}\sum_{\mathbf{k}}(1 - 2n_{C}(\mathbf{k})) |\tilde{V}(\mathbf{k})|.\label{eq:Cc}
   \end{align}
Detailed calculations can be found in Appendix~\ref{ap:caseII}. In the grid model\cite{Boris2004}, the $c-$states are absent, hence, $C_{c}=0$, and the value of $C_{a}$ is reported to be $0.958$.
 
  The same approach as in Case I now gives the critical temperature
\begin{equation}
T_{c}   =  \frac{g^{2}}{4 |\epsilon_{b}|}  \left(\frac{\exp(|\epsilon_{b}|/T_{c})-1}{\exp(|\epsilon_{b}|/T_{c}) + 1} \right) ,
\label{Tc2}
\end{equation}
and the gap parameter 
 \begin{equation}
 \Delta = \frac{g^{2} (1-2n_{B})\tilde{C}}{8\epsilon_{B}}.
 \label{Delta2}
 \end{equation}
associated with the Van Hove singularity for $A$- and $B$- states located at $|V(\mathbf{k})|  = |\tilde{V}(\mathbf{k})| = 2$.

\subsection{Temperature evolution of the energy gap}
\label{evolution}

\begin{figure}
\includegraphics[width=0.48\textwidth]{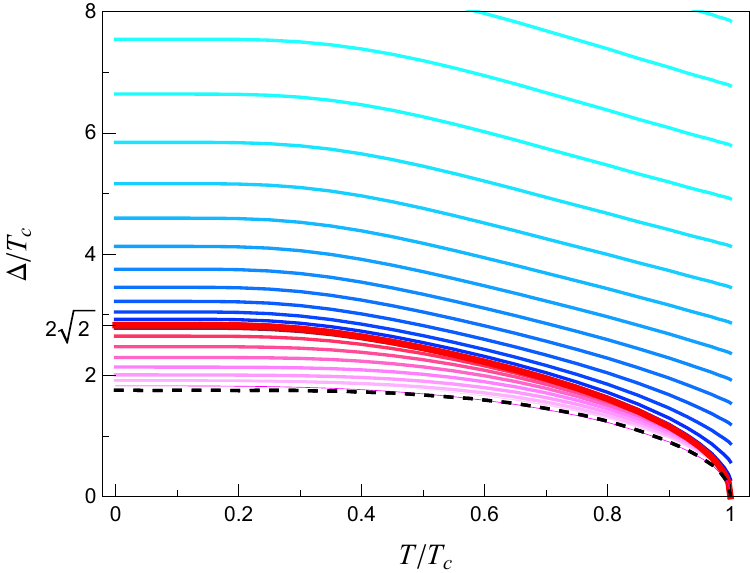}
\caption{(Color online). Family of theoretical curves for  temperature dependence of the superconductong gap for different ratios $\Delta(0)/T_{c}$. Thick red line corresponds to the critical ratio $\Delta(0)/T_{c}=2\sqrt{2}$. Solid lines above  the  thick line represent Case I and below the thick line Case II. The dashed line shows the standard result of the Bardeen-Cooper-Schrieffer theory\cite{BCS1957}.}
\label{fig:Figfamily}
  \end{figure}

In Fig.~\ref{fig:Figfamily}, we present temperature dependences of energy gaps for Cases I and II given by Eqs.(\ref{Delta1}) and (\ref{Delta2}) respectively. These dependencies were obtained by the numerical solution of the system of equations Eqs.(\ref{nB}, \ref{eq:na}, \ref{eq:nc},\ref{eq:EaenergyI}, \ref{eq:EcenergyI}, \ref{eq:EbenergyI}) for Case I, or Eqs.(\ref{nB}, \ref{eq:na}, \ref{eq:nc}, \ref{eq:EbenergyII}, \ref{eq:EaenergyII}, \ref{eq:EcenergyII}) for Case II.
     %%******************************%%  

The families of plots for Cases I and II are connected through the critical case $\epsilon_a = \epsilon_b = \epsilon_c = 0$, which is represented by the thick red line. This case corresponds to the ratio $\Delta(0)/T_c = 2 \sqrt{2} \approx 2.82$. Plots above the critical-case line correspond to Case I: at $T= T_c$, they all end at nonzero values $\Delta(T_c) = \epsilon_a$. Plots below the critical-case line correspond to Case II: they all have $\Delta(T_c) = 0$, and, moreover approach closely the canonical BCS dependence for $\epsilon_b/g \rightarrow \infty$.

Thus, if the assumptions of the present model are valid, the critical-case ratio $\Delta(0)/T_c = 2 \sqrt{2} $ signifies the transition from the conventional behavior $\Delta(T_c) = 0$ for $\Delta(0)/T_c < 2 \sqrt{2}$ to unconventional behavior $\Delta(T_c) \neq 0 $ for $\Delta(0)/T_c > 2 \sqrt{2}$. The value of $\Delta(0)/T_c = 2 \sqrt{2} $ for the critical case makes important quantitative difference from the critical case result $\Delta(0)/T_c = 4 $ for the grid-based model of Refs.\cite{Boris2004,Boris2005}, which involved only {\it a}- and {\it b}-states.   Such a difference was to be expected, because  the coupling between {\it b}- and {\it c}-states in the present model leads to additional energy advantage for the superconducting state and hence higher superconducting transition temperature for the same value of the coupling constant $g$.

\section{Superconducting properties}
\label{superconducting}

\subsection{Anomalous correlation functions}
\label{anomalous}

Bogoliubov transformations (\ref{eq:bgvtransb},\ref{eq:BTaeststes}-\ref{eq:BTcoststes}) can be used to obtain the following anomalous correlation functions for $T< T_c$: 
\begin{eqnarray}
\Psi_a({\mathbf{k}}) &= &
\langle a_{\text{e}}({\mathbf{k}}) a_{\text{o}}(-{\mathbf{k}})
\rangle \nonumber\\
&=& u({\mathbf{k}}) v({\mathbf{k}}) e^{\phi_a({\mathbf{k}})}[2 n_A({\mathbf{k}}) -1],
\label{Psiak}\\
\Psi_c({\mathbf{k}}) &=& 
\langle c_{\text{e}}({\mathbf{k}}) c_{\text{o}}(-{\mathbf{k}})
\rangle \nonumber\\
&=& p({\mathbf{k}}) q({\mathbf{k}}) e^{\phi_c({\mathbf{k}})}[2 n_C({\mathbf{k}}) -1],
\label{Psick}\\
 \Psi_b({\mathbf{r}}_{i\alpha}, {\mathbf{r}}_{j\beta})   &\equiv  & 
\langle b_{i\alpha,-} b_{j\alpha^{\prime},+} \rangle \nonumber\\ 
&=& s w e^{i \varphi_{i\alpha}} (1 - 2 n_B)
\delta({\mathbf{r}}_{i\alpha} - {\mathbf{r}}_{j\beta}),
\label{Psib1}
\end{eqnarray}
where ${\mathbf{r}}_{i\alpha}$ is the position of $\alpha$th {\it b}-state in the {\it i}th unitcell, and $\delta({\mathbf{r}}_{i\alpha} - {\mathbf{r}}_{j\beta})$  is defined as  
Kronecker delta on the discrete
superlattice. Two different components
of the SC order parameter corresponding to {\it a}, {\it c}, 
and {\it b}-states are the correlation functions (\ref{Psiak}), (\ref{Psick}), and (\ref{Psib1}) respectively. 

The anomalous averages for {\it a} and {\it c}-components, in real space, can be written as following,
 \begin{eqnarray}
\Psi_{a}(\mathbf{r}_{\text{e}_{[i,\alpha]}}, \mathbf{r}_{\text{o}_{[j,\beta]}}) 
&\equiv &
\langle a(\mathbf{r}_{\text{e}_{[i,\alpha]}}) a(\mathbf{r}_{\text{o}_{[j,\beta]}}) \rangle,\label{Psiarrdef}\\
\Psi_{c}(\mathbf{r}^{\prime}_{\text{e}_{[i,\alpha]}}, \mathbf{r}^{\prime}_{\text{o}_{[j,\beta]}}) 
&\equiv &
\langle c(\mathbf{r}^{\prime}_{\text{e}_{[i,\alpha]}}) c(\mathbf{r}^{\prime}_{\text{o}_{[j,\beta]}}) \rangle,
\label{Psicrrdef}
\end{eqnarray}
where $\mathbf{r}_{\text{e}_{[i,\alpha]}}$ and $\mathbf{r}_{\text{o}_{[j,\beta]}}$ are, respectively, the positions of an even ($\eta=1$) and odd ($\eta=-1$) {\it a}-states  adjacent to the $b$-states labelled by indices $[i,\alpha]$ and $[j,\beta]$, and, likewise, $\mathbf{r}^{\prime}_{\text{e}_{[i,\alpha]}}$ and
$\mathbf{r}^{\prime}_{\text{o}_{[j,\beta]}}$ are, respectively, the positions of an even ($\zeta=1$) and odd ($\zeta=-1$) {\it c}-states  adjacent to the same $b$-states.
The anomalous averages given by Eqs.\eqref{Psiarrdef} and \eqref{Psicrrdef} have non-zero values only when their two arguments correspond to states of  different kind  (i.e. even and odd).
They can be expressed as
\begin{eqnarray}
\Psi_{a}(\mathbf{r}_{\text{e}}, \mathbf{r}_{\text{o}}) 
&=& \frac{2}{N}\sum_{\mathbf{k}} 
\Psi_{a}({\mathbf{k}}) 
e^{i {\mathbf{k}} \left(\mathbf{r}_{\text{e}} - \mathbf{r}_{\text{o}}\right)},
\label{Psiarr}\\
\Psi_{c}(\mathbf{r}^{\prime}_{\text{e}}, \mathbf{r}^{\prime}_{\text{o}}) 
&= &\frac{2}{N}\sum_{\mathbf{k}} 
\Psi_{c}({\mathbf{k}}) 
e^{i {\mathbf{k}} \left(\mathbf{r}^{\prime}_{\text{e}} - \mathbf{r}^{\prime}_{\text{o}}\right)},
\label{Psicrr}
\end{eqnarray}
where $\Psi_a({\mathbf{k}})$ and $\Psi_c({\mathbf{k}})$ are given by Eq.(\ref{Psiak}) and \eqref{Psick}.
These anomalous averages also obey the following relations: 
\begin{eqnarray}
\Psi_{a}(\mathbf{r}_{\text{e}}, \mathbf{r}_{\text{o}})
&=& - \Psi_{a}(\mathbf{r}_{\text{o}}, \mathbf{r}_{\text{e}}),
\label{Psiarr-}\\
\Psi_{c}(\mathbf{r}^{\prime}_{\text{e}}, \mathbf{r}^{\prime}_{\text{o}})
&=& - \Psi_{c}(\mathbf{r}^{\prime}_{\text{o}}, \mathbf{r}^{\prime}_{\text{e}}),\label{Psicrr-}\\
\Psi_{a}(\mathbf{r}_{\text{e}}, \tilde{\mathbf{r}}_{\text{e}})
&=& \Psi_{a}(\mathbf{r}_{\text{o}}, \tilde{\mathbf{r}}_{\text{o}}) = 0,
\label{Psiarr0}\\
\Psi_{c}(\mathbf{r}^{\prime}_{\text{e}}, \tilde{\mathbf{r}}^{\prime}_{\text{e}})
&=& \Psi_{c}(\mathbf{r}^{\prime}_{\text{o}}, \tilde{\mathbf{r}}^{\prime}_{\text{o}}) = 0.
\label{Psicrr0}
\end{eqnarray}
Relations (\ref{Psiarr-}) and (\ref{Psicrr-})  follow from the
fermionic anticommutation rule. Variables  $\tilde{\mathbf{r}}_{\text{e}}$, $\tilde{\mathbf{r}}_{\text{o}}$, $\tilde{\mathbf{r}}^{\prime}_{\text{e}}$ and $\tilde{\mathbf{r}}^{\prime}_{\text{o}}$  in Eqs.~(\ref{Psiarr0}) and (\ref{Psicrr})represent different positions of the same kind as $\mathbf{r}_{\text{e}}$, $\mathbf{r}_{\text{o}}$, $\mathbf{r}^{\prime}_{\text{e}}$ and $\mathbf{r}^{\prime}_{\text{o}}$ respectively.

The coherence length of the order parameters 
$\Psi_{a}( \mathbf{r}_{\text{e}}$, $\mathbf{r}_{\text{o}})$ and $\Psi_{c}(\mathbf{r}_{\text{e}}$, $\mathbf{r}_{\text{o}})$ should  be inversely proportional to the characteristic $\mathbf{k-}$space scale of  $V({\mathbf{k}})$ and $\tilde{V}({\mathbf{k}})$, respectively. 
The examination of Eqs.\eqref{eq:VK} and \eqref{eq:VKT} reveals that this characteristic scale is $\pi/l$, where $l= 8d$ . Therefore, the coherence length associated with $\Psi_{a}( \mathbf{r}_{\text{e}}, \mathbf{r}_{\text{o}})$ and $\Psi_{c}( \mathbf{r}^{\prime}_{\text{e}}, \mathbf{r}^{\prime}_{\text{o}})$ can be estimated as the modulation period $l$.

The coherence length associated with  $\Psi_b$ is equal to
zero, which means that only {\it b}-states located on the same plaquet form coherent pairs.

We now obtain three quantities that characterize the short-range pair correlations:
\begin{eqnarray}
\Psi_{a(i,\alpha)}  &\equiv&  \Psi_a( \mathbf{r}_{\text{e}_{[i,\alpha]}}, \mathbf{r}_{\text{o}_{[i,\alpha]}})
\equiv \langle a_{\text{e}_{[i,\alpha]}} a_{\text{o}_{[i,\alpha]}} \rangle,
\label{Psiaij}\\
\Psi_{c(i,\alpha)}  &\equiv&  \Psi_c( \mathbf{r}^{\prime}_{\text{e}_{[i,\alpha]}}, \mathbf{r}^{\prime}_{\text{o}_{[i,\alpha]}})
\equiv \langle c_{\text{e}_{[i,\alpha]}} c_{\text{o}_{[i,\alpha]}} \rangle,
\label{Psicij}\\
\Psi_{b(i\alpha)} &\equiv&  \Psi_b({\mathbf{r}}_{i\alpha}, {\mathbf{r}}_{i\alpha}) 
\equiv \langle b_{i\alpha,\downarrow} b_{i\alpha,\uparrow} \rangle.
\label{Psibij}
\end{eqnarray}

In Case~I, the explicit expression for $\Psi_{b(i\alpha)}$ 
can be obtained by substituting
the values of  $s$ and $w$ given by Eqs.\eqref{eq:sw} into Eq.(\ref{Psib1}) 
for ${\mathbf{r}}_{i\alpha} = {\mathbf{r}}_{j\beta}$,
which gives
\begin{equation}
\Psi_{b(i\alpha)}  = 
\frac{1}{2} e^{i \varphi_{i\alpha}} (2 n_B - 1).
\label{Psib1I}
\end{equation}
One can then obtain both $\Psi_{a(i,\alpha)} $ and $\Psi_{c(i,\alpha)} $ by making use of the fact that
\begin{eqnarray}
\Psi_{b(i\alpha)}^* \Psi_{a(i,\alpha)}  &=&\frac {E_{\hbox{\footnotesize int}}^{a}} { 4 g N }
\label{Einta}\\
\Psi_{b(i\alpha)}^* \Psi_{c(i,\alpha)}  &=&\frac {E_{\hbox{\footnotesize int}}^{c}}{4 g N }
\label{Eintc}
\end{eqnarray}
where $E_{\hbox{\footnotesize int}}^{a}$ and $E_{\hbox{\footnotesize int}}^{c}$  are the interaction parts
of the energy \eqref{eq:energyfinal}: 
\begin{multline}
E_{\hbox{\footnotesize int}}^{a} =  2gsw (1- 2n_{B})\\ \times 
   \sum_{\mathbf{k}} u(\mathbf{k})v(\mathbf{k})(1-2n_{A}(\mathbf{k})) |V(\mathbf{k})|, 
   \end{multline}
   \begin{multline}
E_{\hbox{\footnotesize int}}^{c}= 2gsw (1- 2n_{B}) \\ \times
   \sum_{\mathbf{k}}   p(\mathbf{k})q(\mathbf{k})(1-2n_{C}(\mathbf{k}))]|\tilde{V}(\mathbf{k})|.  
\end{multline} 
After $E_{\hbox{\footnotesize int}}^{a}$ and $E_{\hbox{\footnotesize int}}^{c}$ are evaluated with the help of Eqs.(\ref{eq:uvkI}-\ref{eq:TcI}),
one can use Eqs.(\ref{Psib1I},\ref{Einta},\ref{Eintc}) to obtain:
\begin{multline}
\Psi_{a(i,\alpha)}  = \frac{g (1 - 2 n_B)  e^{i \varphi_{i\alpha}} }{8 N} \\ \times
\sum_{{\mathbf{k}}} \frac { [1 - 2 n_A({\mathbf{k}})] |V({\mathbf{k}})|^2 
}{ \varepsilon_A({\mathbf{k}})},
\label{Psia1I}
\end{multline}

\begin{multline}
\Psi_{c(i,\alpha)}  = \frac{g (1 - 2 n_B)  e^{i \varphi_{i\alpha}} } {8 N} \\ \times 
\sum_{{\mathbf{k}}}  \frac{ [1 - 2 n_C({\mathbf{k}})] |\tilde{V}({\mathbf{k}})|^2 
}{\varepsilon_C({\mathbf{k}})}.
\label{Psic1I}
\end{multline}

In Case II, the expressions analogous to (\ref{Psia1I},\ref{Psic1I},\ref{Psib1I}) are
\begin{eqnarray}
\Psi_{a(i,\alpha)}  & = & \frac{1}{4}  e^{i \varphi_{i\alpha}} ,\label{Psia1II}\\
\Psi_{c(i,\alpha)} &  = &\frac{1}{4}  e^{i \varphi_{i\alpha}} ,\label{Psic1II}\\
\Psi_{b(i\alpha)} & = &- \frac{ g \tilde{C}  e^{i \varphi_{i\alpha}} (1 - 2 n_B) }{ 8 \varepsilon_B}.
\label{Psib1II}
\end{eqnarray}

\subsection{Superfluid phase stiffness and emergence of superconductivity}
\label{sec:stiff}

The model introduced in Section~\ref{microscopic} is defined in terms of localized fermionic states with a Hamiltonian that does not contain hopping terms or interaction terms  changing the center of mass of the particles. As a result, the mean-field solution obtained in the preceding sections has zero superfluid stiffness and hence cannot sustain superconductivity\cite{erratum-sf}. 
In order to illustrate this, let us consider the following phase transformation of fermionic operators:
\begin{align} \label{eq:phase_grad_1}
\tilde{a}(\mathbf{r}) & =a(\mathbf{r})e^{-i\theta(\mathbf{r})}  & \tilde{a}^{\dagger}(\mathbf{r}) & = a^{\dagger}(\mathbf{r})e^{i\theta(\mathbf{r})} & 
\\\
\tilde{b}_{\sigma}(\mathbf{r}) & =b_{\sigma}(\mathbf{r})e^{-i\theta(\mathbf{r})} & \tilde{b}^{\dagger}_{\sigma}(\mathbf{r}) & = b^{\dagger}_{\sigma}(\mathbf{r})e^{i\theta(\mathbf{r})} &
\\\
\tilde{c}(\mathbf{r}) & =c(\mathbf{r})e^{-i\theta(\mathbf{r})} & \tilde{c}^{\dagger}(\mathbf{r}) & = c^{\dagger}(\mathbf{r})e^{i\theta(\mathbf{r})} \label{eq:phase_grad_3},
\end{align}
where $\theta(\mathbf{r})$ is the position-dependent quantum phase, subscript $\sigma$ represents the spins of the $b$-states. All other former subscripts ``$i$",``e",``o", and ``$\alpha$" are unambiguously determined by the position $\mathbf{r}$.
Technically, the absence of the superfluid stiffness originates from the fact that the Hamiltonian of our model is invariant under    transformation~(\ref{eq:phase_grad_1}-\ref{eq:phase_grad_3}) with \begin{equation}
    \theta(\mathbf{r}) = \mathbf{G} \cdot \mathbf{r},
    \label{theta}
\end{equation} 
where $\mathbf{G}$ is an arbitrary vector. Therefore, there exists a continuous set of mean-field solutions of equal energy, which can be obtained from each other by applying the above set of transformations.

Despite having zero superfluid stiffness, the mean-field solution of the model based on Hamiltonian~(\ref{eq:originalHamlt}) massively suppresses phase fluctuations of the pair correlations that eventually lead to superconductivity. This happens for two reasons. First, the solution fixes the relative phase of the fermionic pairs occupying $a$- and $c$-states with respect to the $b$-states. Second, each $a$- or $c$-state participates in a fermionic pair with four adjacent states of the same kind but opposite spins, which implies that the phase fluctuations other than the linear ones having form (\ref{theta}) across the entire system do cost energy. The above constraints leave the simultaneous phase fluctuations of $a$-, $b$- and $c$-states  with $\theta(\mathbf{r})$ given by Eq.(\ref{theta})  as a rather narrow channel, through which superconductivity is suppressed. As we illustrate below, this channel is removed and the superconductivity is recovered once arbitrarily small hopping terms are added to the model.

In general, the hopping terms can connect two nearest regions ($a\leftrightarrow b$ and $c\leftrightarrow b$), two next-to-nearest regions ($a\leftrightarrow c$ or $b\leftrightarrow b$), two next-to-next-nearest regions, etc.
Because our goal in this section is to describe the main qualitative effects of hopping, we would like to consider the terms whose impact is simplest to calculate.
Thus we include into consideration the hopping terms between two nearest $a$-states with the same spins and between two nearest $c$-states with the same spins: 
\begin{equation}\label{eq:H_new}
    H'=H+H_{t}
\end{equation}
where
\begin{multline}\label{eq:hopping}
H_{t} = t\sum_{\mathbf{r}_a,\alpha}  a^{\dagger}(\mathbf{r}_a+\mathbf{R}_{\alpha}+\mathbf{R}_{\alpha+1})a(\mathbf{r}_a)
\\
+ t\sum_{\mathbf{r}_c,\alpha}  c^{\dagger}(\mathbf{r}_c+\mathbf{R}_{\alpha}+\mathbf{R}_{\alpha+1})c(\mathbf{r}_c)
   \end{multline}
Here $t$ is the hopping parameter, and $\mathbf{r}_a$ and $\mathbf{r}_c$ run over all $a$- and $c$-plaquets respectively.

For the model with Hamiltonian (\ref{eq:H_new}), the variational mean-field solution can be constructed in the same way as in the preceding sections --- with only one difference, namely, the energies of unpaired fermions occupying $a$- and $c$-states now form a band with the $\mathbf{k}$-dependence:
\begin{equation} \label{epsilonk}
    \mathcal{E}(\mathbf{k})\equiv \epsilon_a+2t\left[ \cos(2k_{x}L)+ \cos(2k_{y}L) \right]
\end{equation}
Here we again assume $\epsilon_a = \epsilon_c$. The Bogoliubov transformations are still defined by Eqs.(\ref{eq:bgvtransb},\ref{eq:BTaeststes},\ref{eq:BTaoststes},\ref{eq:BTceststes},\ref{eq:BTcoststes}).
As shown in Appendix \ref{ap:hopping_solution}, for small $t$, the $\mathbf{k}$-dependence of bare energies (\ref{epsilonk}) changes the values of the Bogoliubov coefficients $u(\mathbf{k})$, $v(\mathbf{k})$, $p(\mathbf{k})$ and $q(\mathbf{k})$ only very little. The variational wave function of the system can still be expressed as \begin{multline} \label{Psi}
    |\Psi\rangle= \prod_{\mathbf{k}} \left( u(\mathbf{k})- v(\mathbf{k})e^{i\phi_a}(\mathbf{k})a_{o}^{\dagger}(-\mathbf{k})a_{e}^{\dagger}(\mathbf{k})   \right)\cdot
    \\ \cdot \left( p(\mathbf{k})- q(\mathbf{k})e^{i\phi_c}(\mathbf{k})c_{o}^{\dagger}(-\mathbf{k})c_{e}^{\dagger}(\mathbf{k})   \right)\cdot
    \\ \cdot \prod_{i\alpha}(s-we^{i\varphi_{i\alpha}}b^{\dagger}_{i\alpha\downarrow}b^{\dagger}_{i\alpha\uparrow})|vac \rangle, 
\end{multline}
and the resulting equation for the critical temperature changes also only slightly. 
We note, however, that the hopping part of Hamiltonian $H_{t}$ introduces additional constraints on phases $\varphi_{i}$,  which are important for computing the superfluid stiffness. These constraints  are derived in Appendix \ref{ap:phases}.

In order to calculate the phase stiffness of the above mean-field solution, we introduce an auxiliary wave function $|\tilde{\Psi}\rangle$ obtained from $|\Psi\rangle$ with the help of transformation (\ref{eq:phase_grad_1}-\ref{eq:phase_grad_3}). Namely, in the right-hand-side of Eq.(\ref{Psi}), we use the inverse Fourier transform for $a$- and $c$-operators defined by Eqs.(\ref{eq:FT}-\ref{eq:FT4}), then replace all lattice subscripts with the position argument $\mathbf{r}$, and, finally, substitute the creation operators $a^{\dagger}(\mathbf{r})$, $b^{\dagger}_{\sigma}(\mathbf{r})$, and $c^{\dagger}(\mathbf{r})$ with their respective counterparts $\tilde{a}^{\dagger}(\mathbf{r})$, $\tilde{b}^{\dagger}_{\sigma}(\mathbf{r})$, and $\tilde{c}^{\dagger}(\mathbf{r})$ defined by Eqs.(\ref{eq:phase_grad_1}-\ref{eq:phase_grad_3}).  
The state represented by the wave function $|\tilde{\Psi}\rangle$ carries supercurrent. It is the lowest-energy state of the Hamiltonian $\tilde{H}'$ which has the same form as $H'$ but in terms of $\tilde{a}(\mathbf{r})$, $\tilde{b}_{\sigma}(\mathbf{r})$ and $\tilde{c}(\mathbf{r})$.

We define the zero-temperature phase stiffness $J$ in the following way:
\begin{equation}\label{eq:J_definition}
    \sum_{\mathbf{r}} \frac{J}{2}(\mathbf{\nabla} \theta)^{2} 4 L^2  =\langle \tilde{\Psi}|H'|\tilde{\Psi}\rangle-\langle \Psi|H'|\Psi\rangle+o((\mathbf{\nabla} \theta)^{2}),
\end{equation}
where $\theta$ is the phase appearing in transformations (\ref{eq:phase_grad_1}-\ref{eq:phase_grad_3}), and the sum in the left-hand side is taken over all $2 L \times 2L$ unit cells of the spin vortex checkerboard. (In the continuum limit, this sum is replaced by $\displaystyle \int \frac{J}{2}(\mathbf{\nabla} \theta)^{2} d^2 {\mathbf{r}}  $.)

In Appendix~\ref{ap:stiffness}, we obtain the following expression for the value of  $J$ corresponding to the Hamiltonian (\ref{eq:H_new}) in the case where $\epsilon_b=\mu\equiv0$:
\begin{multline}
 J= \frac{2t}{N}\sum_{\mathbf{k}}\cos(2k_{x}L) \times \sign{\big(\mathcal{E(\mathbf{k})}\big)} \times
 \\
 \times \left( \sqrt{\frac{\mathcal{E}^2(\mathbf{k})}{\mathcal{E}^2(\mathbf{k})+\frac{1}{4}g^2|V(\mathbf{k})|^2}}\right. +
\\ 
 +\left. \sqrt{\frac{\mathcal{E}^2(\mathbf{k})}{\mathcal{E}^2(\mathbf{k})+\frac{1}{4}g^2|\tilde{V}(\mathbf{k})|^2}} \right)
\end{multline}

We explicitly evaluated this expression in Case I under additional assumption $|t| \ll |g| \ll |\epsilon_a|$. In the leading order, 
the result is: 

\noindent For $\displaystyle \frac{4|t\epsilon_a|}{ g^2}\leqslant 1 $,
\begin{equation}\label{eq:J_result_1}
 J=\frac{4t^{2}}{|\epsilon_a|},
\end{equation}
while, for $\displaystyle \frac{4|t\epsilon_a|}{g^2}\geqslant 1$,
\begin{equation}\label{eq:J_result_2}
J=\frac{|t| g^2}{\epsilon_a^{2}}.
\end{equation}

In this work, we do not attempt to calculate the superconducting properties at finite temperatures. Below we only discuss qualitative considerations about the resulting thermodynamic behavior:

Without hopping, the phase transition that we obtained in the mean-field approximation for Hamiltonian~(\ref{eq:originalHamlt}) turns into a crossover, whose sharpness is limited by the phase fluctuations of the order parameter. 
The addition of a small hopping part (\ref{eq:hopping}) to Hamiltonian~(\ref{eq:originalHamlt})  introduces into the system finite superfluid stiffness and the ability to carry current at sufficiently low temperatures. In such a setting, we expect the temperature $T_{ph}$, at which in-plane superfluid stiffness appears, to be significantly smaller than the the mean-field transition temperature $T_c$. Strong pair fluctuations  should thus be present in the temperature range between $T_{ph}$ and $T_c$. However, in a three-dimensional setting of stacked layers and in the presence of Coulomb interaction between charges, the above fluctuations should be somewhat suppressed.

Once the overall phase of the superconducting solution is stabilized, the strong spatial dependence of the anomalous correlation functions described in Section~\ref{anomalous} can lead to an unorthodox interpretation of the experiments sensitive to the sign of the superconducting phase in cuprates --- see the discussion in Ref.\cite{Boris2004}. It also  implies that the standard factorization of the Cooper-pair wave functions in terms of the center-of-mass dependence and the relative-coordinate dependence is not applicable, which also means that the classification of Cooper pairs in terms of spin singlets and spin triplets does not apply. This in turn, may lead to non-trivial spin susceptibility of the superconducting state, which requires a separate investigation extending beyond the scope of the present paper. Here, we would only like to remark that the current experimental knowledge of the spin susceptibility of superconducting cuprates is largely based on the Knight shift nuclear magnetic resonance (NMR) experiments\cite{Slichter2006}, which are subject to a number of assumptions about the chemical shifts and about certain accidental cancellation of hyperfine coefficients. According to a recent review\cite{Haase2017},  the overall Knight-shift phenomenology in cuprates is still evolving, and this may lead to a re-examination of the standard NMR interpretations adopted in the past.

\section{Theoretical scenarios for cuprates}
\label{scenarios}

\subsection{1/8-doped lanthanum cuprates}
\label{doped}

The model presented in this article is primarily relevant to the properties of 1/8-doped lanthanum cuprates where the static spin and charge order is stabilized, such as La$_{1.875}$Ba$_{0.125}$CuO$_4$ (1/8-doped LBCO). On cooling, the latter system exhibits static charge order at temperatures below $T_{CO}=54$K. On further cooling at zero magnetic field, 1/8-doped LBCO sample investigated in Ref.\cite{Li2007} exhibits the onset of static spin order and a simultaneous sharp drop of in-plane resistivity by a factor of 10 at temperature $T_{SO} \approx 40$K. At yet lower temperature, the in-plane resistivity decays to unmeasurably small values following the Berezinskii-Kosterlitz-Thouless-like \cite{Berezinskii1971,Kosterlitz1973} scaling with characteristic temperature $T_{BKT}=16$K. Finally, at temperature $T_{c}=4$K, the sample exhibits bulk superconductivity. Another sample reported by the same group in Ref.~\cite{li2018tuning} exhibits similar behavior as far as the initial resistivity drop and the bulk superconducting transitions are concerned, but, in the temperature range between 4K and 40K, the temperature dependence of resistivity appears to be sample-dependent. The latter non-universality, probably, indicates a ``fragile'' quantum mechanism of resistivity in the interval between the temperature of the initial resistivity drop and the temperature of the onset of bulk superconductivity.

 A theoretical interpretation of the above phenomenology in the framework of the stripe scenario was presented in Refs.\cite{Berg2007,Berg2009}. It was based on the idea of strongly suppressed Josephson coupling between the adjacent CuO$_2$ layers caused by the in-plane change of the sign of the SC order parameter between the adjacent stripes and then by the mismatch of that sign between the adjacent layers. 
 
\begin{figure}
\includegraphics[width=0.47\textwidth]{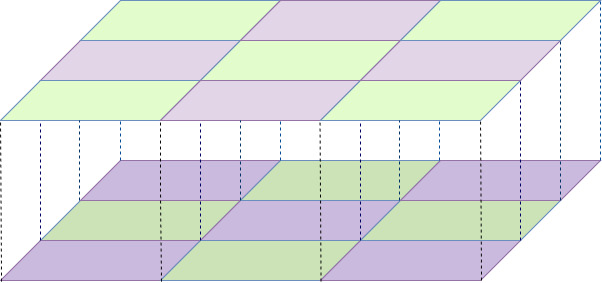}
\caption{(Color online) Two checkerboard patterns in the adjacent planes. Violet squares schematically denote plaquets, corresponding to b-states, and green squares denote antiferromagnetic a or c plaquets.}
\label{fig:stacked}
  \end{figure}

 A similar suppression of the Josephson coupling between adjacent layers can exist in the checkerboard scenario. As shown in Ref.\cite{Boris2004} for the grid checkerboard and in the present work for the spin-vortex checkerboard, the sign of the SC order parameters of the $a$- (and $c$-) fermionic components is supposed to change in real space, remaining zero on average - thus falling under the definition of the pair-density wave proposed in Ref.\cite{Berg2009}. The sign of SC order for the $b$-component may also change in real space but it may or may not lead the zero real-space average. If layers described by our model are stacked on the top of each other, then, due to the Coulomb repulsion, the $b$-plaquets are supposed to be located over the $a$- or $c$-plaquets - see Fig.\ref{fig:stacked}. This mismatch suppresses the Josephson tunneling between the adjacent planes, which can lead to different transition temperatures for the onsets of the two-dimensional and three-dimensional superconductivity.  (We note here that the above-mentioned mismatch cannot be perfect, because the adjacent CuO$_2$ planes in lanthanum cuprates are shifted by half a crystal period.)
 
Apart from the difference between stripes and checkerboards, another  crucial difference between our proposal and that of Ref.\cite{Berg2009} is that the model of Ref.\cite{Berg2009} is based on one fermionic component forming Cooper pairs inside non-magnetic regions, while our model involves two fermionic components --- one residing in the magnetic regions ($a$-/$c$-states) and the other one in non-magnetic regions ($b$-states), and, moreover, the Hamiltonian of our model is dominated by the center-of-mass-conserving interaction between the two components, which leads to anomalously low in-plane superfluid stiffness. 
As discussed at the end of the preceding section, our model is then supposed to exhibit a higher-temperature crossover associated with the appearance of anomalous fermionic averages (\ref{Psib1})-(\ref{Psicrrdef}) with fluctuating phases and the lower-temperature transition that leads to establishing the long-range phase coherence.
We associate the above higher-temperature crossover with the initial resistivity drop in 1/8-doped LBCO at $T \sim 40$K, while the lower-temperature transition is assigned to the observed BKT-like crossover.

Now we address several issues that naturally arise in the context of the above scenario:

(i) {\it Two-dimensional modeling versus three-dimensional character of real materials.} 
In the present work, we only obtain a mean-field description of a two-dimensional model. On the one hand, it is known that long-range fluctuations and the proliferation of topological defects turn the transitions obtained from such a solution into a crossover. On the other hand, the CuO$_2$ planes described by our model are stacked in a three-dimensional setting and, as a result, coupled to each other via Coulomb interaction, lattice strain, spin exchange and the hopping of charge carriers. This, in turn, suppresses both the long-range fluctuations and the formation of topological defects, thereby, making the mean-field solution adequate at least semi-quantitatively.   

(ii) {\it Reason for the resistivity drop to coincide with the mean-field transition temperature in our model.}  Let us consider Case IA of our model in the limit $g \ll \epsilon_a = \epsilon_c$ with small hopping between $a$-states and between $c$-states, as also done in Section~\ref{sec:stiff}. (In terms of the concentration of charge carriers, Case IB with $g \ll |\epsilon_a| = |\epsilon_c|$  is, likely, closer to the situation in 1/8-doped LBCO, but the mathematics of this case is the same as that of Case IA, while the latter is more intuitive to discuss.) In Case IA, $T_c \ll \epsilon_a$ and, therefore, just above $T_c$, the occupations of the $a$- and $c$-states are  exponentially small, which means large resistivity. Below $T_c$, the anomalous averages of $a$- and $c$- states (\ref{Psiarrdef},\ref{Psicrrdef}) appear, which means that Cooper pairs can now occupy these states. Due to anomalously low superfluid stiffness the phase of these Cooper pairs fluctuates. Hence, instead of carrying a supercurrent, they carry normal current but the resulting state still has resistivity much smaller than that above $T_c$. 

(iii) {\it Coincidence between the mean-field transition temperature in our model and the onset of static spin modulations.} In Ref.\cite{Li2007}, the drop of resistivity in zero magnetic field was observed to coincide with the onset static spin modulations. Since we attribute the same resistivity drop to the mean-field transition in our model, the question arises: what can our model have to do with the onset of static spin modulations? In the model, we do not obtain the spin background but rather treat it as imposed externally.  We can imagine, however, that, in a more complete theory, spin-background degrees of freedom can be included in the modeling together with the terms considered by us. In such a case, the energy gain due to the onset of superconducting correlations can simultaneously become a factor stabilizing the spin background, thereby making it static. We would then expect that the magnetic field that suppresses the onset of superconducting fluctuations  should simultaneously suppress the onset of static spin order. Such a behavior can be contrasted with the weak sensitivity to magnetic fields of spin orders driven by the short-range exchange interactions.

\begin{figure*}
\includegraphics[width=0.98\textwidth]{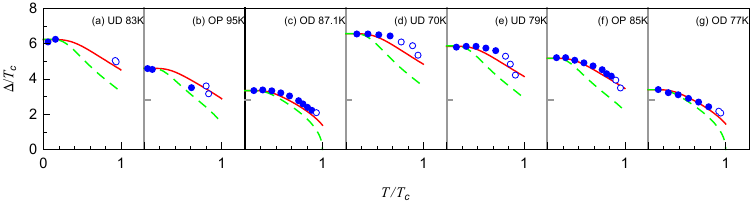}
\includegraphics[width=0.98\textwidth]{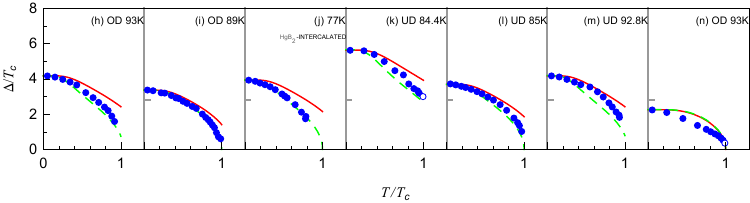}
\includegraphics[width=0.98\textwidth]{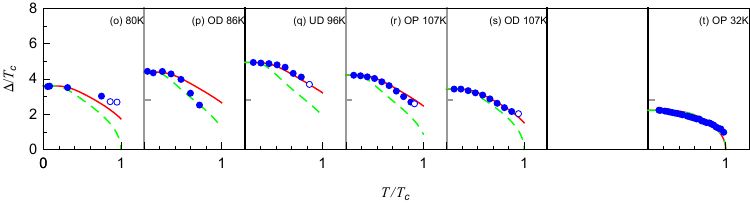}
\caption{(Color online). Temperature evolution of superconducting gap $\Delta(T)$ - part 1 (reviewed in Ref.\cite{Boris2005}). Circles represent experimental data sets for break junction (BJ) and interlayer tunnelling (ILT). [Open circles imply that the data points correspond to very broad and small SC peaks.] Solid red line represents theoretical results of the current work, green dashed line previous theoretical work\cite{Boris2005}.  The experimental data sets are taken from the following references-- (a,b) BJ - Miyakawa {\it et} {\it al.,} \cite{Miyakawa1998}; (c) ILT - Suzuki {\it et} {\it al.,}\cite{Suzuki1999}; (d-g)  ILT - Suzuki and Watanabe \cite{Suzuki2000};  (h,i) ILT- Krasnov {\it et} {\it al.,}\cite{Krasnov2000};  (j) ILT- Krasnov \cite{Krasnov2002}; (k-n) ILT- Krasnov,\cite{Krasnov2002C}; (o) BJ- Vedeneev {\it et} {\it al.,} \cite{Vedeneev1994}; (p) BJ- Akimenko {\it et} {\it al.,} \cite{Akimenko1999}  [plots (a-p) are for Bi-2212];  (q-s) ILT for Bi-2223 from  Yamada {\it et} {\it al.,} \cite{Yamada2003}; and  (t) ILT for Bi-2201-La$_{0.4}$ from Yurgens {\it et} {\it al.} \cite{Yurgens2003}. UD refers to underdoped samples, OP- optimally doped, OD - overdoped. The superconducting critical temperature also indicated in each frame. Horizontal marks in each frame indicate the critical ratio $\Delta/T_{c}=2\sqrt{2}.$ }
\label{fig:exp}
  \end{figure*}

 \begin{figure*}
\includegraphics[width=0.98\textwidth]{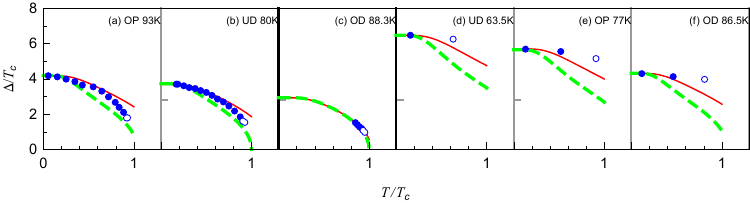}
\includegraphics[width=0.98\textwidth]{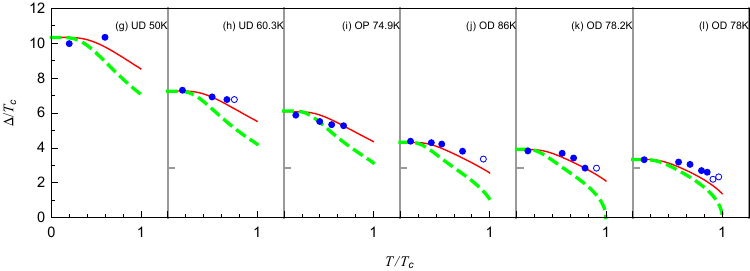}
\includegraphics[width=0.98\textwidth]{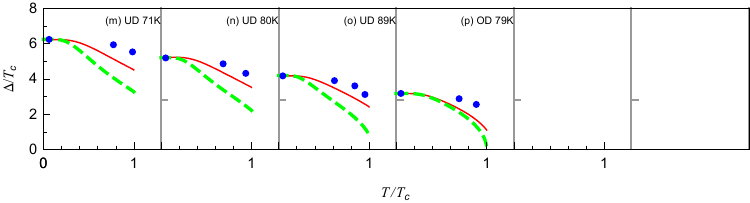}
\caption{(Color online) Temperature evolution of superconducting gap $\Delta(T)$ - part 2 (experiments after 2005). Circles represent experimental data sets for  interlayer tunnelling (ILT). [Open circles imply that the data points correspond to very broad and small SC peaks.] Solid red line represents theoretical results of the current work, green dashed line previous theoretical work\cite{Boris2005}.  The experimental data sets are taken from the following references-- (a,b)  Krasnov \cite{krasnov16}; (c) Bae {\it et} {\it al.,}\cite{park08}; (d,e,f) Kambara {\it et} {\it al.,}\cite{suzuki13}; (g-l) Suzuki {\it et} {\it al.,}\cite{watanabe12}; and (m-p) Ren {\it et} {\it al.,}\cite{zhao12}. All plots except (d,e,f) are for Bi$_{2}$Sr$_{2}$CaCu$_{2}$O$_{8+\delta}$; (d,e,f) are for Bi$_{1.9}$Pb$_{0.1}$Sr$_{2}$CaCu$_{2}$O$_{8+\delta}$.}  
  \label{fig:newexp}
  \end{figure*}

\subsection{Other cuprates}
\label{other}

Let us now turn to cuprates beyond doping 1/8 and beyond the lanthanum family. As shown in Ref.\cite{Fine2008}, cuprates are generically close to the threshold of Coulomb-frustrated phase separation into nanoregions of stronger antiferromagnetic correlations and lower density of charge carriers and nearly non-magnetic nanoregions attracting charge carriers. It is, therefore, reasonable to assume that nanoscale charge inhomogeneities are generically present in cuprates, but they are not necessarily completely static or periodic. (Yet the evidence for charge modulations that are both static and periodic was reported for a variety of cuprates\cite{Hoffman2002,Vershinin2004,Hanaguri2004,Wise2008,Comin2016,peng2018re}.) It is further plausible that, in the general case, clusters with stronger antiferromagnetic correlations are not magnetically ordered with respect to each other. In fact, each of them might be in a singlet state as far as the total spin is concerned (see Ref.\cite{Lychkovskiy2018}). As mentioned earlier, our model in the present form is lacking variables describing the degrees of freedom of the spin background. 
We expect that, generic situation in cuprates is such that the effective hopping between different nanoregions of inhomogeneous electronic background is larger than that for 1/8-doped lanthanum cuprates. This larger hopping can be sufficient to make $T_{ph}$, $T_c$ to approach each other, but still not large enough to dominate over the principal center-of-mass conserving interaction term appearing in Hamiltonian (\ref{eq:originalHamlt}).
Therefore, apart from suppressing the superconducting phase fluctuations, the model with the dynamic spin background can, quite plausibly, have the variational solution close in its principal features to the one obtained in the present work, in particular, as far as the gap of the fermionic excitations is concerned.

\section{Comparison of model predictions with experimentally measured energy gaps}
\label{experiment}

We now demonstrate that the energy gap obtained in Section~\ref{evolution} from the variational solution of the model based on Hamiltonian (\ref{eq:originalHamlt}) exhibits temperature dependence quite similar to that of the superconducting energy gap experimentally observed in a variety of cuprates.
In Figs.~\ref{fig:exp} and \ref{fig:newexp}, we make comparisons with the available experimental results for break junctions (BJ)\cite{Vedeneev1994,Miyakawa1998,Akimenko1999} and the interlayer tunneling (ILT) \cite{Suzuki1999,Suzuki2000,Krasnov2000,Krasnov2002,Krasnov2002C,Yurgens2003,Yamada2003,
park08,watanabe12,zhao12,suzuki13,krasnov16} for the bismuth family of cuprates. Figure ~\ref{fig:exp} includes the data sets reviewed in Ref.\cite{Boris2005}, while Fig.~\ref{fig:newexp} covers the experiments done later. The predictions of the grid-based model are also plotted in Figs.~\ref{fig:exp} and \ref{fig:newexp}.

The model predictions, when limited to Cases I or II only, require two input parameters $\Delta(0)$ and $T_c$, which help us to determine $g$ and $|\epsilon_a|$ for Case I, or $|\epsilon_b|$ for Case II. The choice between Cases I and II is made on the basis of the ratio $\Delta(0)/T_c$ being larger or smaller than $2 \sqrt{2}$. 

 All plots in Figs.~\ref{fig:exp} and \ref{fig:newexp} demonstrate either very good or satisfactory agreement between the predictions of the present model and the experiment. In comparison with the predictions of the grid-based model, the agreement with experiment has improved overall. It should be remarked here that the experimental data themselves are subject to a number of uncertainties, including, in particular, the overheating effect for the ILT measurements \cite{Suzuki1999,Suzuki2000,Yamada2003,Krasnov2000,Krasnov2002,
 Krasnov2002C,Yurgens2003,Zavaritsky2004,Zavaritsky2004C,Yurgens2004,
 Krasnov2005,Kurter2010,Krasnov2011,Kurter2011,watanabe12}.

We further remark that, despite the significant experimental difficulty of measuring $\Delta(T) $ close to $T=T_c$, the very notion of the existence fo the critical ratio  $\Delta(0)/T_c$ which separates the dependencies ending with  $\Delta(T_c) = 0 $ from those ending with $\Delta(T_c) \neq 0 $ appears to be reasonably supported by experiments, and, moreover, the value $2\sqrt{2}$ for such a critical ratio obtained in this work leads to more consistent predictions than the critical value 4 obtained in Ref.\cite{Boris2005} for the grid-based model.

\begin{figure}
\includegraphics[width=0.47\textwidth]{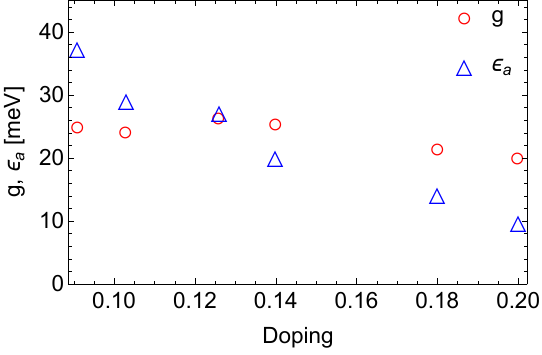}
\caption{(Color online). Doping dependence of the model parameters $g$ and $\epsilon_{a}$ for the calculations presented in Fig.~\ref{fig:newexp} (g-l) to describe ILT experiments of Suzuki {\it et} {\it al.}\cite{watanabe12} for Bi$_{2}$Sr$_{2}$CaCu$_{2}$O$_{8+\delta}$. The model parameters were obtained by solving numerically Eqs.(\ref{Tc1}) and (\ref{Delta1}) with $\epsilon_{b}=0$, $n_B=0$ and with experimentally determined input parameters $\Delta(0)$ and $T_c$. The doping level is as indicated in Ref.\cite{watanabe12}}
\label{fig:gANDea}
  \end{figure}

Finally, in Fig.~\ref{fig:gANDea}, we present the empirical dependence of the model parameters $g$ and $\epsilon_a$ on the doping level. We extract this dependence from the ILT experiments reported in Ref.~\cite{watanabe12}, where the values of doping were explicitly indicated. The parameters $g$ and $\epsilon_a$ were obtained numerically by solving Eqs.(\ref{Tc1}) and (\ref{Delta1}) with $\epsilon_{b}=0$, $n_B=0$ and with experimentally determined input parameters $\Delta(0)$ and $T_c$. The resulting plot shows that, in the doping range between 0.09 and 0.2,  $g$ depends on the doping level rather weakly, while $\epsilon_a$ depends strongly --- it decreases nearly linearly with increasing doping and, if extrapolated, reaches zero around the doping level 0.25, i.e. close to the level, above which superconductivity disappears. Such a behavior of $\epsilon_a$ is consistent with the behavior of the pseudogap expected from other experiments. The eventual disappearance of superconductivity at higher dopings implies, that, if the present model is relevant, then the interaction constant $g$ should steeply decrease to zero for doping levels beyond the plotting range of Fig.~\ref{fig:gANDea} (i.e. above 0.2).

 \section{Conclusions}
 
 In the present paper, we generalized the microscopic model proposed in Ref. \cite{Boris2004} for the grid background to the background formed by the checkerboard of spin vortices. The technical difference is that the former involves two kinds of fermionic states, while the later involves three, even though two of the three are similar.
 We have shown that the predictions of the grid-based model for the temperature evolution of the energy gap largely remain intact. The most important difference between the spin-vortices-based model and the grid-based model turns out to be the critical ratio $\Delta(0)/T_c$ above which the temperature dependence of the energy gap ends at the value $\Delta(T_c) \neq 0$, which, in turn, is, probably, related to the pseudogap. For spin vortices, this critical value is $2\sqrt{2}$, while, for grid, it is 4. We have further demonstrated that the predictions of the spin-vortices-based model for the temperature evolution of the energy gap  exhibit good agreement with experiments, and, moreover, this agreement is somewhat better than that for the grid-based model. 
 
 An important difference of the present work from that of Ref.\cite{Boris2004} is the treatment of the superfluid stiffness of the resulting solution. We have shown that nonlocal character of the interaction term of the kind used in Ref.\cite{Boris2004} is not sufficient to induce non-zero superfluid stiffness. The latter appears only once arbitrarily small hopping terms are added to the model.   The resulting behavior then implies large superconducting fluctuations of the kind that, possibly, exist in 1/8-doped LBCO.

 In the broader context of cuprate superconductivity, the model considered in this work is still rather oversimplified. However, one can  use it to develop intuition about more realistic settings that must involve the fluctuations of the spin background, as well as other interactions between fermions. 

\acknowledgements

The authors are grateful to J. Haase, M. Jurkutat, J. Kohlrautz and A. V. Rozhkov for discussions. This project was funded by the Skoltech NGP Program (Skoltech-MIT joint project).

\appendix 

 \section{Total energy of the system}
 \label{ap:total}
 
In this Appendix, we elaborate on the derivation steps (ii) and (iii) mentioned after Eq.(\ref{eq:uvpqcons}).

Substituting the canonical transformation for $a$- and $c$-states (\ref{eq:BTaeststes}-\ref{eq:BTcoststes}) in (\ref{eq:FTac}), we obtain
%%%%%%%%%%%%%%%%%%%%%%%%%%%%%%%%%%%%%%%
\begin{multline}\label{eq:energywphase}
 E =  8 \epsilon_{b} N \left[ s^{2} n_{B} + w^{2} (1- n_{B}) \right]  \\  + 2 \epsilon_{a} \sum_{\mathbf{k}} \left\lbrace u^{2}(\mathbf{k})n_{A}(\mathbf{k}) + v^{2}(\mathbf{k})[1-n_{A}(\mathbf{k})] \right\rbrace \\+2 \epsilon_{c} \sum_{\mathbf{k}} \left\lbrace p^{2}(\mathbf{k})n_{C}(\mathbf{k}) + q^{2}(\mathbf{k})[1-n_{C}(\mathbf{k})] \right\rbrace   \\
 + 2gsw \left( 1- 2n_{B} \right) \sum_{\mathbf{k}} \\ \bigg[
     u(\mathbf{k})v(\mathbf{k})(1-2n_{A}(\mathbf{k})) |V(\mathbf{k})| \cos [ \phi_{V}(\mathbf{k})+ \phi_{a}(\mathbf{k})] \bigg. \\ \bigg. + p(\mathbf{k})q(\mathbf{k})(1-2n_{C}(\mathbf{k}))|\tilde{V}(\mathbf{k})| \cos [\phi_{\tilde{V}}(\mathbf{k})+ \phi_{c}(\mathbf{k})]\bigg],
    \end{multline}
 where all variables are defined in Section (III).
 
 Two interaction terms in the above expression have phase-dependent factors $\cos [ \phi_{V}(\mathbf{k})+ \phi_{a}(\mathbf{k})]$ and $\cos [\phi_{\tilde{V}}(\mathbf{k})+ \phi_{c}(\mathbf{k})]$. Eventually, the variational ground-state energy obtained by finding $u(\mathbf{k})$, $v(\mathbf{k})$, $p(\mathbf{k})$ and $q(\mathbf{k})$ will monotonically decrease with the increasing absolute value of these terms. This implies that the variational energy will be minimized for $|\cos [ \phi_{V}(\mathbf{k})+ \phi_{a}(\mathbf{k})]|= 1$ and $|\cos [\phi_{\tilde{V}}(\mathbf{k})+ \phi_{c}(\mathbf{k})]|=1$. Choosing the sign of cosines in these relations is just a matter of sign convention for the Bogoliubov transformation coefficients later converting into the sign of the products $sw u(\mathbf{k}) v(\mathbf{k})$ and $sw p(\mathbf{k}) q(\mathbf{k})$. We adopt the convention that the signs of the above products are negative in the state minimizing the total energy.

   \section{Case I}\label{ap:caseI}
  
  For $\epsilon_{b}=0$, the Bogoliubov transformation parameters $s$ and $w$ enter the energy (\ref{eq:energyfinal}) only as a term proportional to $sw$. For such a case, given the constraint $s^2 + w^2 = 1$, the minimization of energy (\ref{eq:energyfinal})  gives, $s  =  \sqrt{\frac{1}{2}},$ $w  =  -\sqrt{\frac{1}{2}}$. The relative negative sign of $s$ and $w$ implies later the positive relative sign for the  pairs of transformation parameters $\{u(\mathbf{k}), v(\mathbf{k})  \}$ and $\{ p(\mathbf{k}), q(\mathbf{k}) \}$. For $\epsilon_{a} \geq 0$, the minimization of energy  (\ref{eq:energyfinal}) with respect to $u(\mathbf{k})$, $v(\mathbf{k}) $, $p(\mathbf{k})$, and $q(\mathbf{k})$, finally, gives
     \begin{eqnarray}\label{eq:uvkI}
    \begin{aligned}
   u(\mathbf{k}) & = & \sqrt{\frac{1}{2}+\frac{1}{2}\sqrt{\frac{1}{1+\frac{\mathcal{T}_{a}^{2}(\mathbf{k})}{\mathcal{Q}_{a}^{2}(\mathbf{k})}}}} ,\\
   v(\mathbf{k}) & = & \sqrt{\frac{1}{2}-\frac{1}{2}\sqrt{\frac{1}{1+\frac{\mathcal{T}_{a}^{2}(\mathbf{k})}{\mathcal{Q}_{a}^{2}(\mathbf{k})}}}},
   \end{aligned}
    \end{eqnarray}
     \begin{eqnarray}
    \begin{aligned}\label{eq:pqkI}
    p(\mathbf{k}) & = & \sqrt{\frac{1}{2}+\frac{1}{2}\sqrt{\frac{1}{1+\frac{ \mathcal{T}_{c}^{2}(\mathbf{k})}{ \mathcal{Q}_{c}^{2}(\mathbf{k})}}}}, \\
   q(\mathbf{k}) & = & \sqrt{\frac{1}{2}-\frac{1}{2}\sqrt{\frac{1}{1+\frac{ \mathcal{T}_{c}^{2}(\mathbf{k})}{ \mathcal{Q}_{c}^{2}(\mathbf{k})}}}}, 
   \end{aligned}
    \end{eqnarray}
      where,
 \begin{eqnarray}\label{eq:TaI}
 \begin{aligned}
  \mathcal{T}_{a}(\mathbf{k})& = g (1-2n_{B}) (1-2n_{A}(\mathbf{k}))|V(\mathbf{k})| ,\\
  \mathcal{Q}_{a}(\mathbf{k})& =  2 \epsilon_{a} (1-2n_{A}(\mathbf{k})),
  \end{aligned}   
   \end{eqnarray}
   \begin{eqnarray}\label{eq:TcI}
    \begin{aligned}
     \mathcal{T}_{c}(\mathbf{k})& = g (1-2n_{B}) (1-2n_{C}(\mathbf{k}))|\tilde{V}(\mathbf{k})|, \\
   \mathcal{Q}_{c}(\mathbf{k})& =  2 \epsilon_{c} (1-2n_{C}(\mathbf{k})).\\
\end{aligned}   
   \end{eqnarray}

The total energy $E$ of the system in this case can be expressed as
    \begin{multline}\label{eq:TenergyI}
    E= -\sum_{k}\bigg[\left(1-2n_{A}(\mathbf{k})\right)\epsilon_{A}(\mathbf{k})\\+\left(1-2n_{C}(\mathbf{k})\right)\epsilon_{C}(\mathbf{k})-\epsilon_{a}-\epsilon_{c}\bigg],
    \end{multline}
which is an implicit function of $|V(\mathbf{k})|$ and $|\tilde{V}(\mathbf{k})|.$ Both $|V(\mathbf{k})|$ and $|\tilde{V}(\mathbf{k})|$ are a function of four phases $\varphi_{1}$, $\varphi_{2}$, $\varphi_{3}$ and $\varphi_{4}$ entering Eqs. (\ref{eq:VK}, \ref{eq:VKT}). Therefore the energy equation should be further minimized with respect to the values of these phases. Such minimization imposes only one constraint\cite{Boris2004}
\begin{equation}\label{eq:phasecondition}
\frac{\varphi_{2}+\varphi_{4}-\varphi_{1}-\varphi_{3}}{2}=\frac{\pi}{2}+\pi n.
\end{equation}

  \section{Case II}\label{ap:caseII}
 In this case, $\epsilon_{a} = \epsilon_{c} = 0 $ in Eq. \eqref{eq:energyfinal} and, as a result, the minimization of energy gives $u(\mathbf{k}) = v(\mathbf{k}) = p(\mathbf{k}) = q(\mathbf{k}) = 1/\sqrt{2}$. The minimization with respect to $s$ subject to condition \eqref{eq:bBog}, now gives
  \begin{equation}
    s^{4} -s^{2} +\frac{\tilde{ \mathcal{T}}^{2}}{4(  \mathcal{Q}^{2}+\tilde{ \mathcal{T}}^{2})}=0
    \label{s4}
   \end{equation}
 where we introduced the following parameters:
  
  \begin{equation}\label{eq:rightcases}
  \begin{rightcases}
    \mathcal{Q}& \equiv  8\epsilon_{b} N (2 n_{B}-1),\\
   \tilde{ \mathcal{T}}& =  \mathcal{T}_{a} +  \mathcal{T}_{c},\\
    \mathcal{T}_{a} &= g(1-2n_{B}) C_{a} N,\\
    \mathcal{T}_{c} &= g(1-2n_{B}) C_{c} N.\\
    \end{rightcases}
     \end{equation}
 The parameters $C_a$ and $C_c$ are defined by Eqs.(\ref{eq:Ca},\ref{eq:Cc}).
 The solution of the biquadratic equation (\ref{s4}) that minimizes the total energy under condition $\epsilon_{b} \geq 0$ is
 \begin{eqnarray}\label{eq:sw}
 \begin{aligned}
   s&=& \sqrt{\frac{1}{2}+\frac{1}{2}\sqrt{\frac{1}{1+\frac{\tilde{ \mathcal{T}}^{2}}{ \mathcal{Q}^{2}}}}},\\
   w&=& -\sqrt{\frac{1}{2}-\frac{1}{2}\sqrt{\frac{1}{1+\frac{\tilde{ \mathcal{T}}^{2}}{ \mathcal{Q}^{2}}}}},
   \end{aligned}
    \end{eqnarray}
      \\ 
 We obtain $\epsilon_{B}$ by varying energy \eqref{eq:energyfinal} with respect to $n_{B}$: 
  \begin{multline*}
  \epsilon_{B} \equiv \frac{1}{8N}\frac{dE}{dn_{B}} =  \\
   \epsilon_{b}  [ s^{2}  -w^{2}]  + 2gsw (-2) \sum_{\mathbf{k}}   \left\{ u(\mathbf{k})v(\mathbf{k})(1-2n_{A}(\mathbf{k})) |V(\mathbf{k})|  \right. \\ 
   \left . + p(\mathbf{k})q(\mathbf{k})(1-2n_{C}(\mathbf{k}))(\mathbf{k})]|\tilde{V}(\mathbf{k})| \right\}.
    \end{multline*}
 The substitution of the parameters defined by Eqs.(\ref{eq:sw},\ref{eq:rightcases})  now gives   
     \begin{equation*}
    \epsilon_{B} =   \epsilon_{b}  \sqrt{\frac{ \mathcal{Q}^{2}}{ \mathcal{Q}^{2}+\tilde{ \mathcal{T}}^{2}}} - 2\frac{\tilde{ \mathcal{T}}}{2\sqrt{ \mathcal{Q}^{2}  +\tilde{ \mathcal{T}}^{2}}}\frac{1}{(1-2n_{B})} \tilde{ \mathcal{T}},
    \end{equation*}
  which, after some manipulations, leads to Eq.(\ref{eq:EbenergyII}). 
 
 The quasiparticle excitation energies for $A$- and $C$- states are calculated in a similar way --- as $\epsilon_{A}(\mathbf{k})\equiv \frac{1}{2}\frac{dE}{dn_{A}(\mathbf{k})}$ and $\epsilon_{C}(\mathbf{k}) \equiv \frac{1}{2} \frac{dE}{dn_{C}(\mathbf{k})}$, which gives Eqs.(\ref{eq:EaenergyII}) and (\ref{eq:EcenergyII}) respectively.  The total energy of the system thus becomes  
 \begin{equation}\label{eq:TenergyII}
    E = -2N[\left(1-2n_{B}\right)\epsilon_{B}-\epsilon_{b}].
    \end{equation}
In this case, phases $\varphi_{\alpha}$ also obey the constraint (\ref{eq:phasecondition}).

\section{Effects of non-zero hopping on the variational solution}\label{ap:hopping_solution}

In this Appendix, we discuss the properties of the superconducting variational solution for Hamiltonian given by Eq.~(\ref{eq:H_new}).
We introduce Bogoliubov transformations in exactly the same way as was done in Section \ref{solution}. Then the energy of the system has the following form:

\begin{multline} \label{eq:energy_with_hopping} 
 E =  8 \epsilon_{b} N \left[ s^{2} n_{B} + w^{2} (1- n_{B}) \right]  \\+ 2 \sum_{\mathbf{k}} \mathcal{E}({\mathbf{k}}) \left\lbrace u^{2}(\mathbf{k})n_{A}(\mathbf{k}) + v^{2}(\mathbf{k})[1-n_{A}(\mathbf{k})] \right\rbrace \\+2 \sum_{\mathbf{k}} \mathcal{E}({\mathbf{k}}) \left\lbrace p^{2}(\mathbf{k})n_{C}(\mathbf{k}) + q^{2}(\mathbf{k})[1-n_{C}(\mathbf{k})] \right\rbrace \\+ 2gsw \left( 1- 2n_{B} \right) 
   \sum_{\mathbf{k}} \bigg[ u(\mathbf{k})v(\mathbf{k})(1-2n_{A}(\mathbf{k})) |V(\mathbf{k})|\bigg. + \\ \bigg.+ p(\mathbf{k})q(\mathbf{k})(1-2n_{C}(\mathbf{k}))]|\tilde{V}(\mathbf{k})| \bigg],
  \end{multline}
where $\mathcal{E}(\mathbf{k})$ is the band energy given by equation (\ref{epsilonk}). The only aspect that is different here in comparison with Eq.(\ref{eq:energyfinal}) is the $\mathbf{k}$-dependence of the band energies for the $a$- and $c$-states. 

We are unable to find an analytic expression for the Bogoliubov transformation coefficients in the general case, but we still can do it in Case I ($\epsilon_b$=0), where the minimization of energy (\ref{eq:energy_with_hopping}) gives  $s  =  \sqrt{\frac{1}{2}},$ $w  =  -\sqrt{\frac{1}{2}}$. The relative negative sign of $s$ and $w$ implies later the positive relative sign for the  pairs of transformation parameters $\{u(\mathbf{k}), v(\mathbf{k})  \}$ and $\{ p(\mathbf{k}), q(\mathbf{k}) \}$. The minimization of energy  (\ref{eq:energy_with_hopping}) with respect to $u(\mathbf{k})$, $v(\mathbf{k}) $, $p(\mathbf{k})$, and $q(\mathbf{k})$, leads to the result having the same form as the one given by equations (\ref{eq:uvkI})-(\ref{eq:pqkI}), with the only difference being that the parameters $\mathcal{T}_{a}(\mathbf{k})$, $\mathcal{Q}_{a}(\mathbf{k})$, $\mathcal{T}_{c}(\mathbf{k})$ and $\mathcal{Q}_{c}(\mathbf{k})$ are now defined as:
 \begin{eqnarray}
 \label{TQa2}
 \begin{aligned}
  \mathcal{T}_{a}(\mathbf{k})& = g (1-2n_{B}) (1-2n_{A}(\mathbf{k}))|V(\mathbf{k})| ,\\
  \mathcal{Q}_{a}(\mathbf{k})& =  2 \mathcal{E}(\mathbf{k}) (1-2n_{A}(\mathbf{k})),
  \end{aligned}   
   \end{eqnarray}
   \begin{eqnarray}
   \label{TQc2}
    \begin{aligned}
    \mathcal{T}_{c}(\mathbf{k})& = g (1-2n_{B}) (1-2n_{C}(\mathbf{k}))|\tilde{V}(\mathbf{k})|, \\
  \mathcal{Q}_{c}(\mathbf{k})& =  2 \mathcal{E}(\mathbf{k}) (1-2n_{C}(\mathbf{k})).\\
\end{aligned}   
   \end{eqnarray}
The distinction between Eqs.(\ref{TQa2},\ref{TQc2}) and Eqs.(\ref{eq:TaI}, \ref{eq:TcI}) is that the former contain the band energy $\mathcal{E}(\mathbf{k})$ given by Eq.(\ref{epsilonk}), while the latter contain the on-site energies $\epsilon_a$ and $\epsilon_c$. Equation (\ref{Tc1}) for the critical temperature is now replaced by the following one:
 \begin{multline}\label{eq:Tc_hopping}
 T_{c}   =  \frac{g^{2}}{32N}  \sum_{\mathbf{k}} \Bigg[ \frac{\big( \exp{(|\mathcal{E}(\mathbf{k})|/T_{c})}-1 \big)}{\big( \exp{(|\mathcal{E}(\mathbf{k})|/T_{c})}+1 \big)} 
 \\
 \times \frac{1}{|\mathcal{E}(\mathbf{k})|} \big( |V(\mathbf{k})|^2+ |\tilde{V}(\mathbf{k})|^2\big) \Bigg]
\end{multline}

\section{Additional constraints on phases $\varphi_{i}$ caused by hopping}\label{ap:phases}
 
Energy minimization for Hamiltonian (\ref{eq:H_new}) leads to additional constraints on phases $\varphi_{\alpha}$ appearing in Bogoliubov transformations (\ref{eq:bgvtransb}). In this Appendix, we derive these constraints for the case $\epsilon_b=0$.

For convenience, let us introduce   new phases:
\begin{equation}
\alpha \equiv \frac{\varphi_{1}-\varphi_{2}-\varphi_{3}+\varphi_{4}}{2} ,
\end{equation}
\begin{equation}
\beta \equiv \frac{\varphi_{1}+\varphi_{2}-\varphi_{3}-\varphi_{4}}{2} ,
\end{equation}
\begin{equation}
\gamma \equiv \frac{-\varphi_{1}+\varphi_{2}-\varphi_{3}+\varphi_{4}}{2} ,
\end{equation}
and then express $|V(\mathbf{k})|^2$ and $|\tilde{V}(\mathbf{k})|^2$ from Eqs. (\ref{eq:VK}), (\ref{eq:VKT}) as
\begin{multline}\label{Vabg1}
     |V(\mathbf{k})|^2=  4\Big( 1 +\cos{(2k_{x}L+\alpha)}\cdot\cos{(2k_{y}L+\beta)} +
     \\
     +\cos{\gamma} \cdot \big[\cos{(2k_{x}L+\alpha)} + \cos{(2k_{y}L+\beta)} \big]\Big)
\end{multline}
and
\begin{multline}\label{Vabg2}
     |\tilde{V}(\mathbf{k})|^2=  4\Big( 1 +\cos{(2k_{x}L+\alpha)}\cdot\cos{(2k_{y}L-\beta)} +
     \\
     +\cos{\gamma} \cdot \big[\cos{(2k_{x}L+\alpha)} + \cos{(2k_{y}L-\beta)} \big]\Big).
\end{multline}

In the presence of non-zero hopping $t$, the expression (\ref{eq:TenergyI}) for the total energy of the system is modified as follows:
\begin{multline}
        E=-\sum_{\mathbf{k}} \left[ \left( \sqrt{\mathcal{E}^{2}(\mathbf{k})+\frac{1}{4}g^{2}|V(\mathbf{k})|^{2}} -\mathcal{E}(\mathbf{k}) \right) + \right.
        \\
        +\left. \left( \sqrt{\mathcal{E}^{2}(\mathbf{k})+\frac{1}{4}g^{2}|\tilde{V}(\mathbf{k})|^{2}} -\mathcal{E}(\mathbf{k}) \right)  \right],
        \label{Etott}
\end{multline}
where $\mathcal{E}(\mathbf{k})$ is the band energy given by Eq.(\ref{epsilonk}).

Necessary conditions for phases $\alpha$, $\beta$ and $\gamma$ to minimize the total energy (\ref{Etott}) are:
\begin{empheq}[left=\empheqlbrace]{align}
\frac{\partial E}{\partial \alpha} & =0, \label{eq:phi_minimize_E_1} \\
 \frac{\partial E}{\partial \beta} & =0, \label{eq:phi_minimize_E_2}\\
\frac{\partial E}{\partial \gamma} & =0. \label{eq:phi_minimize_E_3} 
\end{empheq}
In general, the system of equations (\ref{eq:phi_minimize_E_1})-(\ref{eq:phi_minimize_E_1}) has more than one solution for a given set of parameters $\epsilon_a$, $g$ and $t$. The solution representing the global minimum should then be selected by computing the corresponding values of $E$.  

We were able to find analytically three types of solutions of the system (\ref{eq:phi_minimize_E_1})-(\ref{eq:phi_minimize_E_3}).

\noindent 
Solutions of type  A:
\begin{equation}
\alpha =\pi n, \beta=\pi m, \gamma=\pi k \label{eq:phases_critical_1}.
\end{equation}
Solutions of type  B:
\begin{equation}
 \alpha =\pi n,  \beta=\pi (n-1)+2\pi m,  \gamma=\pi k + \frac{\pi}{2}.
 \label{eq:phases_critical_2}
\end{equation}
Solutions of type  C (exist, only if $\displaystyle \frac{4|t\epsilon_a|}{ g^2}\leqslant 1 $):
\begin{empheq}{align}
\alpha= 2\pi n,  \beta=2\pi m,  \cos{\gamma}=- \frac{4t\epsilon_{a}}{g^2} ;
\label{eq:phases_critical_3a}  
\\
\alpha= (2 n+1)\pi,  \beta=(2m+1)\pi,  \cos{\gamma}= \frac{4t\epsilon_{a}}{g^2} .
\label{eq:phases_critical_3b}
\end{empheq}
Here  $k$, $m$ and $n$ are arbitrary integer numbers.
 We cannot prove analytically that the system (\ref{eq:phi_minimize_E_1})-(\ref{eq:phi_minimize_E_1}) has no other solutions, but we checked numerically that the minimum of $E$ corresponds to the solutions of either type A, or B, or C. 

Let us now prove that the sets of $\alpha$, $\beta$ and $\gamma$ given by  Eqs.(\ref{eq:phases_critical_1}-\ref{eq:phases_critical_3b}) are, indeed, the solutions of the system (\ref{eq:phi_minimize_E_1})-(\ref{eq:phi_minimize_E_3}).

The substitution of Eq.(\ref{Etott}) into Eqs.(\ref{eq:phi_minimize_E_1}) and (\ref{eq:phi_minimize_E_2}) gives, respectively,
\begin{multline}
 \sum_{\mathbf{k}}\frac{\sin(2k_xL+\alpha)\times \cos(2k_yL+\beta)}{\sqrt{\mathcal{E}^{2}(\mathbf{k})+\frac{1}{4}g^{2}|V(\mathbf{k})|^{2}}}+
 \\
 +\cos{\gamma}\sum_{\mathbf{k}} \frac{\sin(2k_xL+\alpha)}{\sqrt{\mathcal{E}^{2}(\mathbf{k})+\frac{1}{4}g^{2}|V(\mathbf{k})|^{2}}}+
 \\
 +\sum_{\mathbf{k}}\frac{\sin(2k_xL+\alpha)\times \cos(2k_yL-\beta)}{\sqrt{\mathcal{E}^{2}(\mathbf{k})+\frac{1}{4}g^{2}|\tilde{V}(\mathbf{k})|^{2}}}+
 \\
 +\cos{\gamma}\sum_{\mathbf{k}} \frac{\sin(2k_xL+\alpha)}{\sqrt{\mathcal{E}^{2}(\mathbf{k})+\frac{1}{4}g^{2}|\tilde{V}(\mathbf{k})|^{2}}}
 =0,
\label{der1}
\end{multline}
and 
\begin{multline}
      -\sum_{\mathbf{k}}\frac{\cos(2k_{x}L+\alpha)\times \sin(2k_{y}L+\beta)}{\sqrt{\mathcal{E}^{2}(\mathbf{k})+\frac{1}{4}g^{2}|V(\mathbf{k})|^{2}}}-
      \\
      -\cos{\gamma}\sum_{\mathbf{k}} \frac{\sin(2k_{y}L+\beta)}{\sqrt{\mathcal{E}^{2}(\mathbf{k})+\frac{1}{4}g^{2}|V(\mathbf{k})|^{2}}}+
   \\
 +\sum_{\mathbf{k}}\frac{\cos(2k_{x}L+\alpha)\times \sin(2k_{y}L-\beta)}{\sqrt{\mathcal{E}^{2}(\mathbf{k})+\frac{1}{4}g^{2}|\tilde{V}(\mathbf{k})|^{2}}}+
 \\
 +\cos{\gamma}\sum_{\mathbf{k}} \frac{\sin(2k_{y}L-\beta)}{\sqrt{\mathcal{E}^{2}(\mathbf{k})+\frac{1}{4}g^{2}|\tilde{V}(\mathbf{k})|^{2}}}=0.
 \label{der2}
\end{multline}
We now observe that $\alpha=\pi n$, $\beta = \pi m $ and arbitrary $\gamma$ satisfy both Eq.(\ref{der1}) and Eq.(\ref{der2}), because  these equations would then contain only summations  of functions that are odd with respect to either  $k_x$ or $k_y$, while the summation regions are symmetric.  Thus all sets of $\alpha$, $\beta$ and $\gamma$ entering (\ref{eq:phases_critical_1}) -(\ref{eq:phases_critical_3b}) satisfy  Eqs.(\ref{der1}) and (\ref{der2}), and, hence, Eqs.(\ref{eq:phi_minimize_E_1}) and (\ref{eq:phi_minimize_E_2})

%(Note that, in the presence of non-zero hopping, we are not able to shift the boundaries of the Brillouin zone as was done in Ref.~\cite{Boris2004}. As a result, Eqs.(\ref{der1}) and (\ref{der2}) are no longer trivially satisfied).

The substitutions of Eq.(\ref{Etott}) into Eq.(\ref{eq:phi_minimize_E_3}) yields:
\begin{multline}
    \sin{\gamma} \Big(  \sum_{\mathbf{k}}\frac{\cos(2k_{x}L+\alpha) + \cos(2k_{y}L+\beta)}{\sqrt{\mathcal{E}^{2}(\mathbf{k})+\frac{1}{4}g^{2}|V(\mathbf{k})|^{2}}}    +
    \\
    + \sum_{\mathbf{k}}\frac{\cos(2k_{x}L+\alpha) + \cos(2k_{y}L-\beta)}{\sqrt{\mathcal{E}^{2}(\mathbf{k})+\frac{1}{4}g^{2}|\tilde{V}(\mathbf{k})|^{2}}}  \Big)=0.
    \label{der3}
\end{multline}

The solutions of type A given by Eq.(\ref{eq:phases_critical_1}) imply that $\sin{\gamma}=0$. Hence they, obviously, satisfy Eq.(\ref{der3}).

Now let us show that the solutions of type B also satisfy Eq.(\ref{der3}). Once we substitute (\ref{eq:phases_critical_2}) to (\ref{der3}), we get an expression, proportional to 
\begin{equation}
\sum_{\mathbf{k}}\frac{\cos(2k_{x}L) - \cos(2k_{y}L)}{\sqrt{\mathcal{E}^{2}(\mathbf{k})+g^{2}\cos(2k_{x}L)\cos(2k_{y}L)}} .   
\end{equation}
This expression is zero, because it changes sign under the transformation $(k_x,k_y)\longrightarrow(k_y,k_x)$.

\begin{figure}
\includegraphics[width=0.47\textwidth]{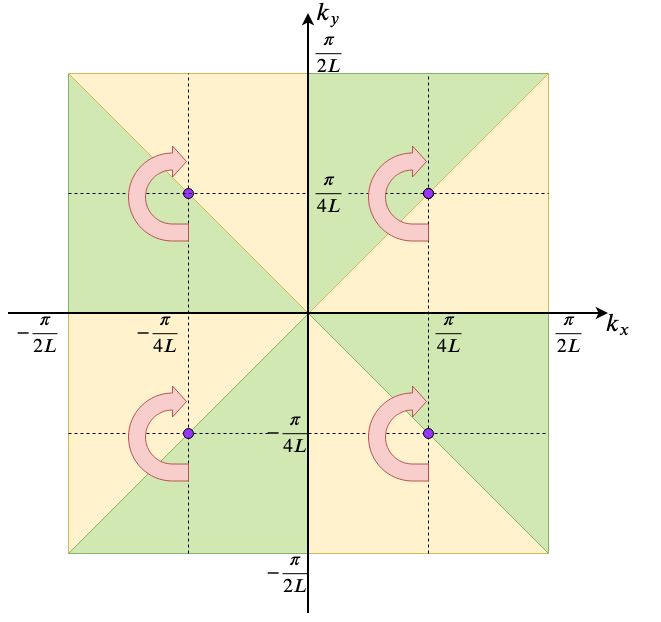}
\caption{(Color online) Schematic representation of transformation (\ref{eq:zone_rotation}) in the Brillouin zone of spin vortex checkerboard.
Each of eight triangle regions is rotated around the nearest purple dot $(\pm\frac{\pi}{4 L},\pm\frac{\pi}{4 L})$ by 180 degrees --- thereby orange regions map into green ones and vice versa.}
\label{fig:brillouin}
  \end{figure}

Finally, let us consider solutions of type C given by (\ref{eq:phases_critical_3a}) and (\ref{eq:phases_critical_3b}) and prove that they  satisfy Eq.(\ref{der3}). In order to do it, let us examine the expressions under the square roots in the denominators in Eq.(\ref{der3}):
\begin{multline}\label{eq:denominator_1}
    \mathcal{E}^{2}(\mathbf{k})+\frac{1}{4}g^{2}|V(\mathbf{k})|^{2}=\epsilon_a^{2}+\\
    + 4t\epsilon_{a} \big[ \cos{\left( 2k_{x}L \right)}+\cos{\left( 2k_{y}L \right)}  \big] +\\
    +4t^2 \big[  \cos{\left( 2k_{x}L \right)}+\cos{\left( 2k_{y}L \right)}  \big]^2+\\
    +g^2\Big( 1 +\cos{(2k_{x}L+\alpha)}\cdot\cos{(2k_{y}L+\beta)} +
     \\
     +\cos{\gamma} \big[\cos{(2k_{x}L+\alpha)} + \cos{(2k_{y}L+\beta)} \big]\Big)
\end{multline}
and
\begin{multline}\label{eq:denominator_2}
    \mathcal{E}^{2}(\mathbf{k})+\frac{1}{4}g^{2}|\tilde{V}(\mathbf{k})|^{2}=\epsilon_a^{2}+\\
    + 4t\epsilon_{a} \big[ \cos{\left( 2k_{x}L \right)}+\cos{\left( 2k_{y}L \right)}  \big] +\\
    +4t^2 \big[  \cos{\left( 2k_{x}L \right)}+\cos{\left( 2k_{y}L \right)}  \big]^2+\\
    +g^2\Big( 1 +\cos{(2k_{x}L+\alpha)}\cdot\cos{(2k_{y}L-\beta)} +
     \\
     +\cos{\gamma} \big[\cos{(2k_{x}L+\alpha)} + \cos{(2k_{y}L-\beta)} \big]\Big).
\end{multline}
When we substitute (\ref{eq:phases_critical_3a}) or (\ref{eq:phases_critical_3b}) into (\ref{eq:denominator_1})-(\ref{eq:denominator_2}), the terms $4t\epsilon_{a} \big[ \cos{\left( 2k_{x}L \right)}+\cos{\left( 2k_{y}L \right)}  \big]$ and $g^2\cos{\gamma} \big[\cos{(2k_{x}L+\alpha)} + \cos{(2k_{y}L\pm\beta)} \big]$ cancel each other.
For this reason, the sum over $\mathbf{k}$ in Eq.(\ref{der3}) for $\alpha$, $\beta$ and $\gamma$ given by (\ref{eq:phases_solution_1}) becomes proportional to 
\begin{multline}\label{eq:sum_zero}
   \sum_{\mathbf{k}} \big[ \cos{\left( 2k_{x}L \right)}+\cos{\left( 2k_{y}L \right)} \big] \times\\
   \times \Big\{ \epsilon^{2}_{a}+4t^{2}\left[\cos{(2k_{x}L)}+\cos{(2k_{y}L)}\right]^{2}+ \\
   +  g^2\big[ 1+\cos{(2k_{x}L)}\cos{(2k_{y}L)} \big] \Big\}^{-\frac{1}{2}}.
  \end{multline}
Now we note that, after the simultaneous change of the signs of $\cos{(2k_xL)}$ and $\cos{(2k_yL)}$, the function under the sum also changes sign. A possible transformation achieving this  is : 
\begin{equation}\label{eq:zone_rotation}
    (k_x,k_y)\longrightarrow \left( \frac{\pi}{2L}\sign(k_x)-k_x,\frac{\pi}{2L}\sign(k_y)-k_y \right)
\end{equation}
It is illustrated in Fig. \ref{fig:brillouin}. The fact that transformation (\ref{eq:zone_rotation}) maps the Brillouin zone onto itself, while the function summed in Eq.(\ref{eq:sum_zero}) changes sign, means that the sum itself is equal to zero and, hence, Eq.(\ref{der3}) is satisfied. This finishes the proof that Eqs.(\ref{eq:phases_critical_1}-\ref{eq:phases_critical_3b}) represent the solutions of the system (\ref{eq:phi_minimize_E_1}-\ref{eq:phi_minimize_E_3}) and thus correspond to the extrema of $E$.

\begin{figure}
\includegraphics[width=0.47\textwidth]{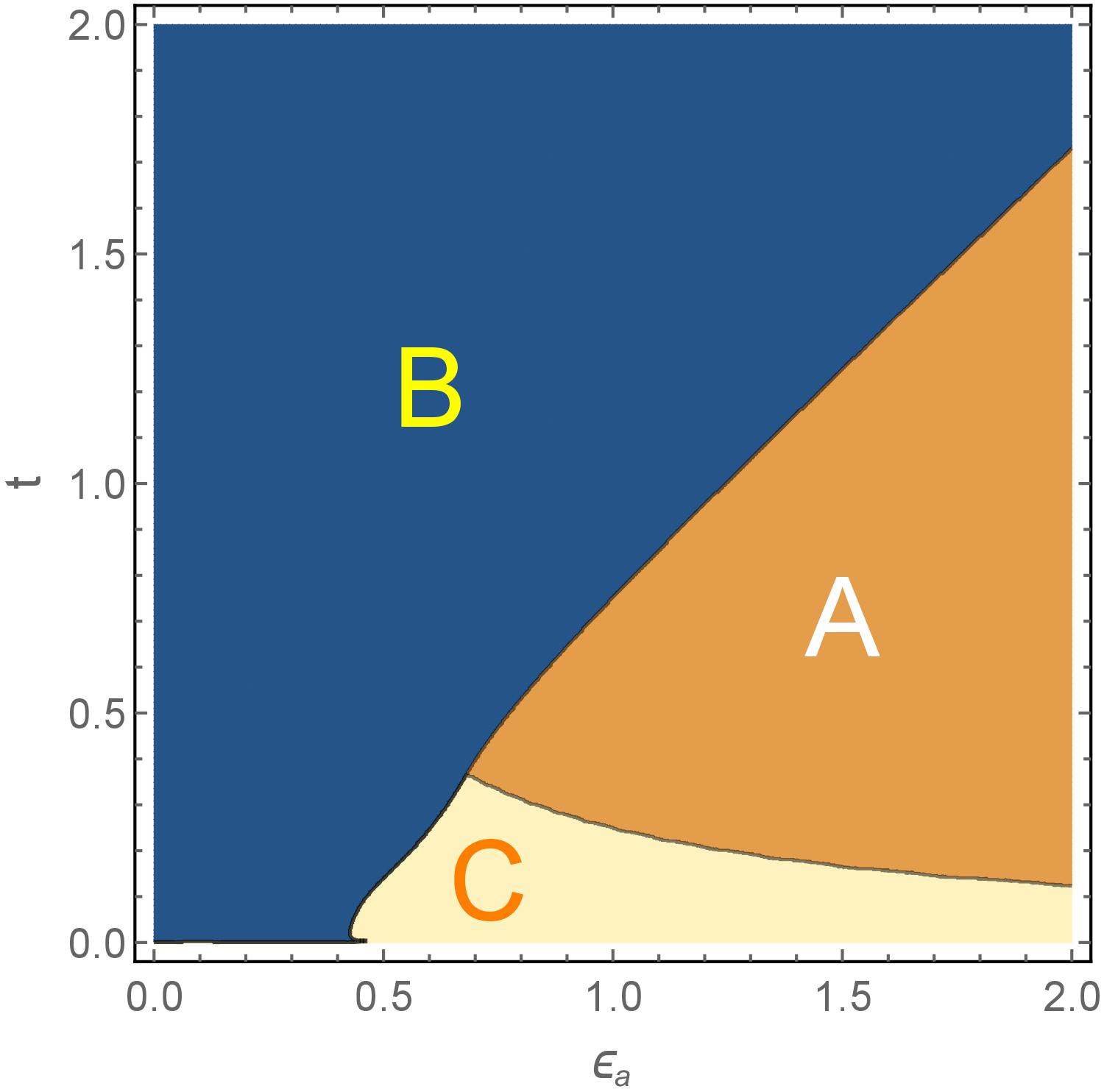}
\caption{(Color online) Numerically obtained  ``phase diagram" of the type of extrema (A,B, or C) given by Eqs.(\ref{eq:phases_critical_1}-\ref{eq:phases_critical_3b}) that realize the global minimum of energy (\ref{Etott}) for $g=1$ and for the values of parameters $(\epsilon_a, t)$ indicated on the axes.}
\label{fig:density_plot}
\end{figure}

We have conducted extensive numerical tests of the minima of E in the space of parameters $\epsilon_a$, $g$ and $t$. In all these tests, the minimuma corresponded to the solutions of type A, B or C given by Eqs.(\ref{eq:phases_critical_1}-\ref{eq:phases_critical_3b}). The numerically found ``phase diagram'' assigning the regions of the parameter space $(\epsilon_a, t)$ to different types of solutions for $g=1$ is shown in Fig. \ref{fig:density_plot}.

In Section~\ref{sec:stiff} and in Appendix~\ref{ap:stiffness}, we focus on the limit
$\displaystyle |t|\ll |g| \ll |\epsilon_a|$. In this case, the minima of $E$ are the following:

\noindent If $\displaystyle \frac{4|t\epsilon_a|}{ g^2}\leqslant 1 $, then
\begin{equation}\label{eq:phases_solution_1}
\left[
  \begin{array}{lcc}
     \cos{\gamma}=-\frac{4t\epsilon_{a}}{g^2}, & \alpha=2\pi n, & \beta=2\pi m;  \\
     \text{or} & & \\
     \cos{\gamma}=\frac{4t\epsilon_{a}}{g^2}, & \alpha=(2n+1)\pi, & \beta=(2m+1)\pi . 
  \end{array}
\right.
 \end{equation}
 
\begin{figure}
\includegraphics[width=0.47\textwidth]{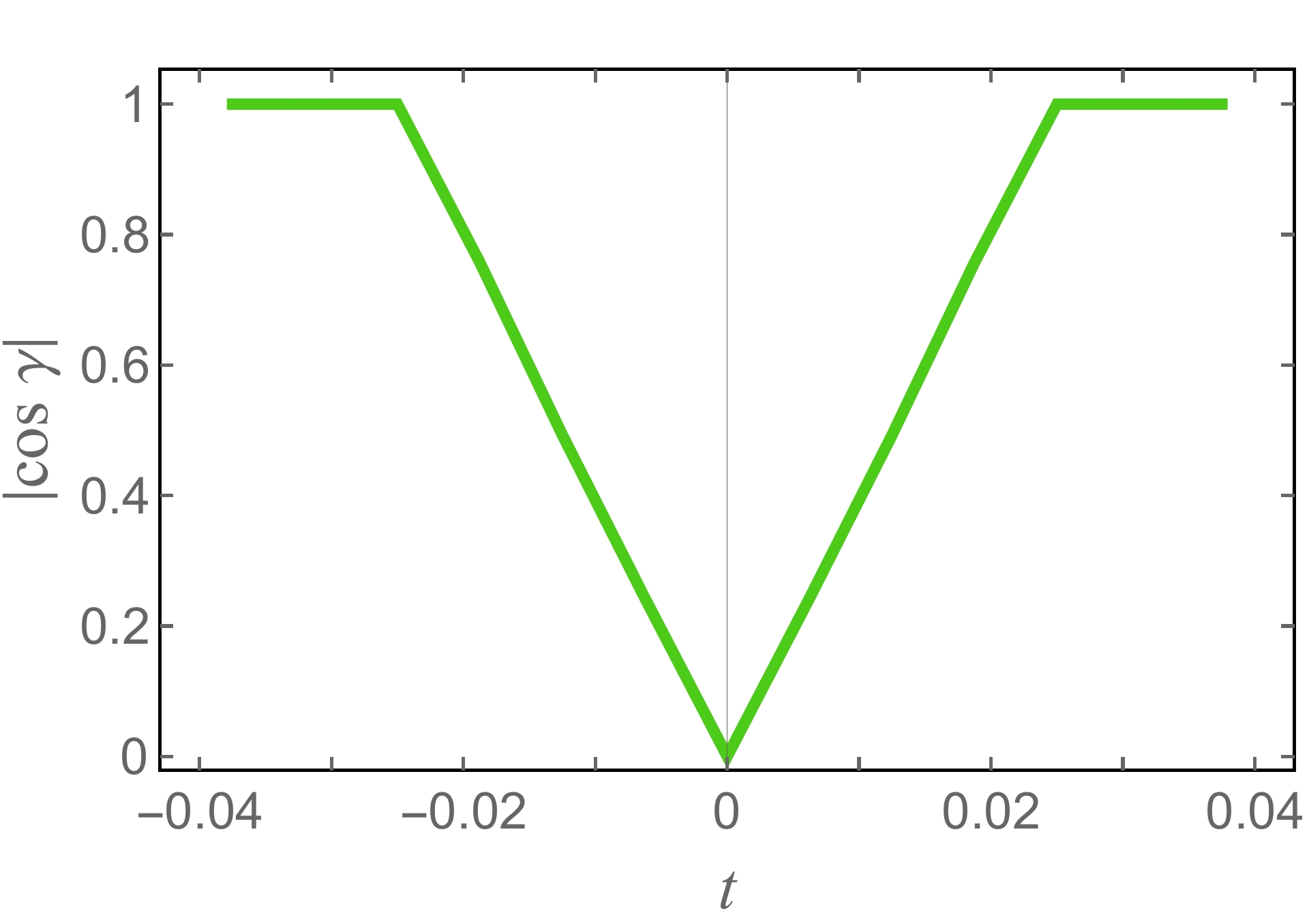}
\caption{(Color online) Numerically obtained $|\cos \gamma \, |$ realizing the global minimum of energy (\ref{Etott}) as a function of $t$ for $g=1$ and $\epsilon_a=10$. It illustrates solutions (\ref{eq:phases_solution_1})-(\ref{eq:phases_solution_2}) in the limit $|t|\ll |g|\ll |\epsilon_a|$: at  $t=\pm\frac{g^2}{4\epsilon_a}= \pm 0.025$ the minimum switches from type C given by Eq.(\ref{eq:phases_solution_1})  to type A given by Eq.(\ref{eq:phases_solution_2}).}
\label{fig:cos_gamma_t}
\end{figure}

\noindent If $\displaystyle \frac{4|t\epsilon_a|}{ g^2}\geqslant 1 $, then
\begin{equation}\label{eq:phases_solution_2}
\left[
  \begin{array}{lcc}
     \cos{\gamma} = - \sign{(t\epsilon_a)}, & \alpha=2\pi n, & \beta=2\pi m  \\
     \text{or} & & \\
     \cos{\gamma} = \sign{(t\epsilon_a)}, & \alpha=(2n+1)\pi, & \beta=(2m+1)\pi . \\
  \end{array}
\right.
 \end{equation}
The transition between the two kinds of minima is illustrated in Fig.~\ref{fig:cos_gamma_t} through the evolution of $|\cos{\gamma}|$ as a function of $t$ for $\epsilon_a = 10$ and $g=1$.

\section{Calculation of superfluid phase stiffness for Hamiltonian (\ref{eq:H_new})}
\label{ap:stiffness}

In this section, we obtain zero-temperature superfluid phase stiffness for Hamiltonian (\ref{eq:H_new}) with $\epsilon_b= 0$.

Let us assume for simplicity that linearly changing in space phase $\theta$ has gradient only along the  $x$-direction. We also introduce variable $\theta_0$ to denote the change of phase $\theta$ across the $2L \times 2L$ unit cell of the spin-vortex checkerboard. The gradient of $\theta$ can now be expressed as 
\begin{equation}
|\mathbf{\nabla}\theta|=\frac{\partial \theta}{\partial x}=\frac{\theta_0}{2L}    
\end{equation}
and, as a result, the energy (\ref{eq:J_definition}) of the system associated with the phase gradient can be written as
\begin{equation}
   \sum_{\mathbf{r}} \frac{J}{2}(\mathbf{\nabla} \theta)^{2} 4 L^2 = \frac{J}{2}N\theta_0^{2}.
\end{equation}
Using Eq.(\ref{eq:J_definition}), we then obtain
\begin{equation}
J=\frac{1}{N} \frac{\partial^2}{\partial \theta_0^2} \Big( \langle \tilde{\Psi}|H'|\tilde{\Psi}\rangle-\langle \Psi|H'|\Psi\rangle \Big)\bigg|_{\theta_0=0}.
\label{JA1}
\end{equation}
We now observe that $\langle \tilde{\Psi}|H'|\tilde{\Psi}\rangle = \langle \Psi|\tilde{H}'|\Psi\rangle $, where  $\tilde{H}'$ is the Hamiltonian obtained by  replacing operators $a$, $b$ and $c$ in Hamiltonian (\ref{eq:H_new}) with operators $\tilde{a}$, $\tilde{b}$ and $\tilde{c}$ defined by Eqs.(\ref{eq:phase_grad_1}-\ref{eq:phase_grad_3}). This allows us to rewrite Eq.(\ref{JA1}) as 
\begin{equation}\label{eq:another_formula_for_J}
 J=\frac{1}{N} \frac{\partial^2}{\partial \theta_0^2} \Big( \langle \Psi|\tilde{H}' - H'|\Psi\rangle \Big)\bigg|_{\theta_0=0}  ,
\end{equation}
where 
\begin{multline}
     \tilde{H}' - H' = \tilde{H}_{t}-H_{t}=
     \\
     =\sum_{\mathbf{k}} 2t[\cos(2k_{x}L+\theta_0)-\cos(2k_{x}L)](a^{+}_{e}(\mathbf{k})a_{e}(\mathbf{k})+\\
     + a^{+}_{o}(\mathbf{k})a_{o}(\mathbf{k})+
     +c^{+}_{e}(\mathbf{k})c_{e}(\mathbf{k})+c^{+}_{o}(\mathbf{k})c_{o}(\mathbf{k}))=\\
     =\sum_{\mathbf{k}}2t[\cos(2k_{x}L)(\cos(\theta_0)-1)-\sin(2k_{x}L)\sin({\theta_0})]\times\\
     \times (a^{+}_{e}(\mathbf{k})a_{e}(\mathbf{k})+a^{+}_{o}(\mathbf{k})a_{o}(\mathbf{k})+c^{+}_{e}(\mathbf{k})c_{e}(\mathbf{k})+c^{+}_{o}(\mathbf{k})c_{o}(\mathbf{k})).
\end{multline}
For the mean-field ground state in the presence of hopping elements, one can use Eqs.(\ref{eq:uvkI}) and (\ref{eq:pqkI}) with ingredients from Eqs.(\ref{TQa2}) and (\ref{TQc2})  to obtain
\begin{multline}\label{average1}
\langle \Psi |a^{+}_{e}(\mathbf{k})a_{e}(\mathbf{k})|\Psi \rangle=\langle \Psi |a^{+}_{o}(\mathbf{k})a_{o}(\mathbf{k})|\Psi \rangle=v^2(\mathbf{k})=\\
=\frac{1}{2} \Bigg( 1-\sign{\big(\mathcal{E}(\mathbf{k})\big)}\sqrt{\frac{\mathcal{E}^2(\mathbf{k})}{\mathcal{E}^2(\mathbf{k})+\frac{1}{4}g^2|V(\mathbf{k})|^2}}\Bigg)
\end{multline}
and
\begin{multline}\label{average2}
\langle \Psi |c^{+}_{e}(\mathbf{k})c_{e}(\mathbf{k})|\Psi \rangle=\langle \Psi |c^{+}_{o}(\mathbf{k})c_{o}(\mathbf{k})|\Psi \rangle=q^2(\mathbf{k})=\\
=\frac{1}{2} \Bigg( 1-\sign{\big(\mathcal{E}(\mathbf{k})\big)}\sqrt{\frac{\mathcal{E}^2(\mathbf{k})}{\mathcal{E}^2(\mathbf{k})+\frac{1}{4}g^2|\tilde{V}(\mathbf{k})|^2}}\Bigg).
\end{multline}

After substituting Eqs.(\ref{average1}) and (\ref{average2}) into Eq.(\ref{eq:another_formula_for_J}), we, finally, get
\begin{multline}\label{Jgeneral}
 J= \frac{2t}{N}\sum_{\mathbf{k}}\cos(2k_{x}L) \times \sign{\big(\mathcal{E(\mathbf{k})}\big)} \times
 \\
 \times \left( \sqrt{\frac{\mathcal{E}^2(\mathbf{k})}{\mathcal{E}^2(\mathbf{k})+\frac{1}{4}g^2|V(\mathbf{k})|^2}}\right. +
\\ 
 +\left. \sqrt{\frac{\mathcal{E}^2(\mathbf{k})}{\mathcal{E}^2(\mathbf{k})+\frac{1}{4}g^2|\tilde{V}(\mathbf{k})|^2}} \right).
\end{multline}
(Here and below, we do not include terms that give zero after the summation.)

Now let us study expression (\ref{Jgeneral}) in the limit $|t| \ll |g| \ll |\epsilon_a|$ and find its dependence on $t$. Expanding it in the powers of $g/\epsilon_a$ and $t/g$, we get, in the leading order, 
 \begin{equation}
J=-\frac{t g^2}{4N\epsilon_a^{2}}\sign{(\epsilon_a)}\sum_{\mathbf{k}}(|V(\mathbf{k})|^{2}+|\tilde{V}(\mathbf{k})|^{2})\cos(2k_{x}L).
 \end{equation}
 
For $\alpha$, $\beta$, $\gamma$ corresponding to the energy minima (\ref{eq:phases_solution_1}) and (\ref{eq:phases_solution_2}), we obtain from Eqs.(\ref{Vabg1}) and (\ref{Vabg2}) that,

\noindent if $\displaystyle \frac{4|t\epsilon_a|}{ g^2}\leqslant 1 $:
\begin{multline}
  |\tilde{V}(\mathbf{k})|^2=|V(\mathbf{k})|^2 = 4\Big( 1 +\cos(2k_xL)\cdot\cos(2k_yL) -\\ -\frac{4t\epsilon_a}{g^2} \big[\cos(2k_xL) + \cos(2k_yL) \big]\Big);
\end{multline}

\noindent and,

\noindent if $\displaystyle \frac{4|t\epsilon_a|}{ g^2}\geqslant 1 $:
\begin{multline}
 |\tilde{V}(\mathbf{k})|^2= |V(\mathbf{k})|^2 = \\
 =4\Big( 1 +\cos(2k_xL)\cdot\cos(2k_yL) - \\
 -\sign(t\epsilon_a)\big[\cos(2k_xL) + \cos(2k_yL) \big]\Big).
\end{multline}
This leads to the following result:

\noindent if $\displaystyle \frac{4|t\epsilon_a|}{ g^2}\leqslant 1 $:
\begin{equation}
 J=\frac{8t^2}{N|\epsilon_a|} \sum_{\mathbf{k}} \cos^{2}{(2k_xL)}=\frac{4t^{2}}{|\epsilon_a|} ;
\end{equation}

\noindent if $\displaystyle \frac{4|t\epsilon_a|}{ g^2}\geqslant 1 $:
\begin{equation}
 J=\sign(\epsilon_{a})\sign(t\epsilon_{a})\frac{2t g^2}{N\epsilon_a^{2}}\cdot \sum_{\mathbf{k}} \cos^{2}{(2k_xL)} =\frac{|t| g^2}{\epsilon_a^{2}} .
\end{equation}

%\bibliography{aa_references}


\end{document}